%% file: main.tex
\documentclass[
reprint,
superscriptaddress,
amsmath,amssymb,
aps,pra,
longbibliography
]{revtex4-1}

\usepackage{graphicx,xcolor,colortbl}

\usepackage{mathtools}
\usepackage{xspace} 

\newcommand{\QuESTlink}{\texttt{QuESTlink}\xspace}

\usepackage{hyperref}
\usepackage{cleveref}

\newcommand{\ie}{\textit{i}.\textit{e}.,~}
\newcommand{\eg}{\textit{e}.\textit{g}.,~}

\usepackage{tikz}
\usetikzlibrary{quantikz}

\usepackage{physics}

\usepackage{bm}

\renewcommand{\vec}[1]{\boldsymbol{\mathrm{#1}}}

\usepackage{booktabs}
\usepackage[export]{adjustbox}
\usepackage[caption=false]{subfig}

\usepackage{array,multirow}
\newcolumntype{C}[1]{>{\centering\arraybackslash}m{#1}}
\newcolumntype{M}[1]{>{\centering\arraybackslash}m{#1}}

\newcommand{\orcid}[1]{\href{https://orcid.org/#1}{\includegraphics[width=8pt]{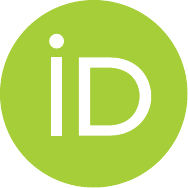}}}

\begin{document}

\title{Exploiting subspace constraints and ab initio variational methods for quantum chemistry}
\author{\orcid{0000-0003-0558-4685}\,Cica Gustiani}
\email{cica.gustiani@materials.ox.ac.uk}
\affiliation{Department of Materials, University of Oxford, Parks Road, Oxford OX1 3PH, United Kingdom}
\author{Richard Meister}
\affiliation{Department of Materials, University of Oxford, Parks Road, Oxford OX1 3PH, United Kingdom}
\author{Simon C. Benjamin}
\email{simon.benjamin@materials.ox.ac.uk}
\affiliation{Department of Materials, University of Oxford, Parks Road, Oxford OX1 3PH, United Kingdom}
\affiliation{Quantum Motion, 9 Sterling Way, London N7 9HJ, United Kingdom}
\date{\today}



\begin{abstract}

Variational methods offer a highly promising route to exploiting quantum computers
for chemistry tasks.
%
%
Here we employ methods described in a sister paper to the present report,
entitled {\it exploring ab initio machine synthesis of quantum circuits}, in order to
solve problems using adaptively evolving quantum circuits. Consistent with
prior authors we find that this approach can outperform human-designed circuits
such as the coupled-cluster or hardware-efficient ans\"atze, and we make
comparisons for larger instances up to $14$ qubits.
Moreover we introduce a novel approach to constraining the circuit
evolution in the physically relevant subspace, finding that this greatly improves 
performance and compactness of the circuits. We consider both static and dynamics 
properties of molecular systems.
%
The emulation environments used is \QuESTlink; all
resources are open source and linked from this paper.
\end{abstract}

\maketitle

 
\section{Introduction}

For over a century, the problem of simulating quantum systems has been a
practical challenge in physics and chemistry. This problem is classically hard
--- even with many approximations --- since the complexity grows exponentially
in space and time.  On the other hand, quantum computers are expected to be capable of tackling such tasks
efficiently in a variety of scenarios. In the current phase of quantum technology, the so-called
NISQ~\cite{preskill2018quantum} era (\emph{noisy intermediate-scale quantum}), we
endeavour to find the first practical examples of such a quantum advantage.

Having a NISQ device means having access to hundreds to thousands of qubits.
This scale is, however, not enough to run fully fault-tolerant computations, thus
limits us to run only shallow circuits. Variational quantum algorithms
(VQAs) are the most widely studied class of algorithms that are potentially compatible with NISQ device contraints. The
VQAs operate on parameterised quantum circuits (ans\"atze) and rely on classical processors for the
parameter evolution, therefore, the quantum circuits
remain shallow. The shared quantum/classical nature of the paradigm means that these approaches are also called `hybrid' quantum algorithms.

An eigensolver -- finding the ground or other eigenstates of a Hamiltonian of interest -- was the first VQA that was proposed nearly a decade ago~\cite{peruzzo2014variational}.   The ground-state energy
problem is provably hard, known for having complexity
QMA-complete~\cite{kempe2004the}. Since then, many variational quantum
algorithms have emerged for various applications, \eg
chemistry~\cite{grimsley2019adaptive,ollitrault2020hardware,delgado2021variational,gibbs2021long,greene2021generalized,metcalf2020resource,chan2021molecular,mcardle2020quantum,li2019variational},
solving
equations~\cite{xu2021variational,endo2020variational,patil2022variational,liu2021variational},
and quantum circuit
compilations~\cite{khatri2019quantum,jones2022robust,gibbs2021long,gokhale2019partial,caro2021generalization}.
We refer to references
\cite{li2019variational,cerezo2021variational,tilly2021variational,fedorov2022vqe}
for overviews and current progress reports on the VQAs.

The choice of ansatz is critical for a successful optimisation, where
it must be expressive enough to approximate the ground state, but not contain
too many parameters as it may cause vanishing gradients~\cite{mcclean2018barren}.  Several
strategies have been proposed to devise a good ansatz, \eg
problem-inspired ans\"atze~\cite{cade2020strategies,martin2021simulating},
hardware-efficient ans\"atze~\cite{kandala2017hardware,benedetti2021hardware,rattew2019domain,tang2021qubit},
unitary coupled cluster ans\"atze~\cite{grimsley2019adaptive,greene2021generalized,sim2021adaptive,tang2021qubit}, and adaptive structure
ans\"atze~\cite{grimsley2019adaptive,tang2021qubit,chivilikhin2020mog,rattew2019domain,chivilikhin2020mog}.
We refer to references \cite{cerezo2021variational,tilly2021variational} for an overview of ansatz
constructions.

Our focus in this work is on techniques where ans\"atze are evolved on the fly
by random guesses followed by the application of selection criteria and judicious deletion. We only consider noise-free
gates, and we performed our algorithms on an emulated quantum computer with up to 14
qubits.  This work is accompanied by our study on various existing and new
optimisation technicalities in the sister paper~\cite{sister}; here, we apply
techniques we found to perform well.

We test our VQA techniques to solve well-known chemistry problems: finding the
ground state of a Hamiltonian and synthesising shallow circuits which implement the
time-evolution operator. For the latter, we consider both full Hilbert space dynamics, as well as circuits created under the more permissive condition that dynamics should be correct in a specific subspace; specifically, this is a in a subspace
that preserves the occupation number. The subspace constrained time-evolution operator (propagator) is
obtained via the \emph{subspace compilation technique} that is introduced
in our sister paper~\cite{sister}.  
Our interest in the subspace-restricted propagators stems from the fact that
most chemical reactions preserve the total number of particles, making
processes outside of this subspace irrelevant. We find that it is more efficient
to employ propagators that operate in a subspace rather than in the entire Hilbert
space. To our knowledge, no prior method to construct such propagators has been explored in the literature.

Our technique is agnostic to the Hamiltonian and can operate without or with only trivial
knowledge of the system. We test it on finding the ground state of small
molecules, using up to 14 qubits. We obtained remarkable gate count reductions
compared to other techniques that we are aware of. For instance, we only need a
total of 20 gates to approximate the ground state of the LiH molecule, which is a
12-qubit problem.  We also test our technique to synthesise the time-evolution
operators of the $\text H_2$ molecule dynamics over various time intervals.
Interestingly, our circuits have nearly-constant gate counts: no more than
18 gates for the subspace restricted compilation and no more than 34 gates 
for the full space compilation; this is considerably fewer than the canonical 
\emph{Trotter-Suzuki} product formulas that need hundreds of gates, even with the optimal choice of Trotter order. Our shallow
propagators can be concatenated and thus potentially pave the way to achieving longer periods of modelled time even in NISQ simulations.

The layout of this paper is the following. Firstly, in
\Cref{sec:chemistry_problems}, we discuss the general formalism and the
optimisation frameworks: \Cref{sec:finding_the} for the ground state estimation
problem and \Cref{sec:the_time-evolution} for the circuit synthesis of the
time-evolution operator constrained to an arbitrary subspace.  Secondly,
\Cref{sec:discussion_on} outlines the optimisation techniques.  Then,
\Cref{sec:results_and} shows all the numeric results together with
discussions and analysis. \Cref{sec:results_and} is divided into two
subsections by the problems: \Cref{sec:res_ground_state} for the ground state
estimation and \Cref{sec:res_the_time-evolution} for the circuit synthesis of
the time-evolution operator. Finally, \Cref{sec:conclusion_and} concludes this
works followed by a general discussion.

\section{Chemistry problems with 2\textsuperscript{nd} quantised Hamiltonians}\label{sec:chemistry_problems}

Estimating and preparing the ground state and
implementation of the time-evolution operator are the
first steps toward dynamical molecular simulation. Here, we tackle these problems 
by means of variational quantum algorithms.

To simulate a molecular Hamiltonian on a quantum computer, one needs to
represent the problem in a format that can be solved on a quantum computer, \eg by
operating quantum gates. In our study, we describe the electronic Hamiltonian
using second quantization, namely 
\begin{equation}
    H = \sum_{i,j}h_{ij}a^\dagger_i a_j + \sum_{i,j,k,l}h_{ijkl}a^\dagger_ia^\dagger_j a_k a_l,   
    \label{eq:Hamiltonian}
\end{equation}
where $\{a_j\}$ and $\{a^\dagger_j\}$ are lowering and raising operators acting
on a Fock space. The coefficients $h_{ij}$ and $h_{ijkl}$ are overlaps and exchange
integrals on the specified basis set; these can be efficiently computed on
classical computers. We use the basis set STO-3G~\cite{szabo2012modern},
namely, contracted three Gaussian functions in which parameters are obtained by
a self-consistent field procedure. 

We proceed by mapping the Fock space onto the Hilbert space of the qubits. We
use the canonical Jordan-Wigner
transformation~\cite{wigner1928paulische,seeley2012bravyi} to map the Fock
states into qubit states in the computational basis and the Fock operators into
quantum gates.
Finally, our Hamiltonian is expressed in the form of a Pauli sum,
\begin{equation}\label{eq:paulisum}
    H=\sum_k c_k P_k,
\end{equation}
where $c_k\in\mathbb R$ and ${P_k}\in\{I,\sigma_x,\sigma_y,\sigma_z\}^{\otimes n}$
are $n$-qubit Pauli strings. 
We utilise the Python packages \texttt{Openfermion}~\cite{mcclean2017openfermion} and \texttt{PyScf}~\cite{sun2018pyscf} to generate our
Hamiltonians. 
After the transformations, our electronic problems are ready to be processed on a quantum computer.  From
this point onwards, we assume the Hamiltonians have the form of
~\Cref{eq:paulisum}.

\subsection{Finding the ground state}\label{sec:finding_the}

Variational quantum algorithms discover the ground state of a Hamiltonian $H$ by 
minimising the energy expectation value (cost function)
\begin{equation}\label{eq:expected}
    \min_{\vec\theta} \bra*{\psi_0}\mathcal C^\dagger(\vec\theta) H\mathcal C(\vec\theta)\ket*{\psi_0},
\end{equation}
where $\mathcal C(\vec\theta)$ is a parameterised quantum circuit (ansatz) and
$\ket*{\psi_0}$ is an initial fixed state. By linearity,
the cost function can be written as 
\begin{equation}\label{eq:expectedsum}
    \sum_k c_k\bra*{\psi_0}\mathcal C^\dagger(\vec\theta) P_k\mathcal C(\vec\theta)\ket*{\psi_0}.
\end{equation}
Quantum computers can efficiently compute each term above independently, 
therefore, the circuits can remain shallow.

When the iterations reach convergence by some criterion, \eg 
cost improvement from the previous iteration is less than $\delta$,
the optimisation halts and returns the optimised parameters $\vec\theta'$.
Then, the resulting ground state energy estimation is 
$E_{gs}\approx^{\varepsilon}\bra*{\psi_0}\mathcal C(\vec\theta')^\dagger H
\mathcal C(\vec\theta')\ket*{\psi_0}$, with the wave function $\mathcal
C(\vec\theta')\ket*{\psi_0}$, and has an error $\varepsilon$. The error
$\varepsilon$ is defined as the difference to the exact ground state energy.
Our goal is for this final energy estimate error $\varepsilon$ to be within the
chemical accuracy $\varepsilon=1.59\times 10^{-3}$ Hartree (1 kcal/mole), which
is the typical error relevant to thermodynamic experiments~\cite{pople1999nobel}.

\subsection{The time-evolution operator in the subspace}\label{sec:the_time-evolution}

The time-evolution operators,
\begin{equation}
    U(\Delta t)=e^{-i\Delta t H},
\end{equation}
are usually hard to compute for interesting problems, \ie when Hamiltonian terms are
not mutually commuting.  The canonical way to obtain $U(\Delta t)$ is 
by employing the Suzuki-Trotter formula~\cite{trotter}; this process is called \emph{Trotterisation}. 

\begin{figure}[h]
    \centering
    \includegraphics[width=.9\columnwidth]{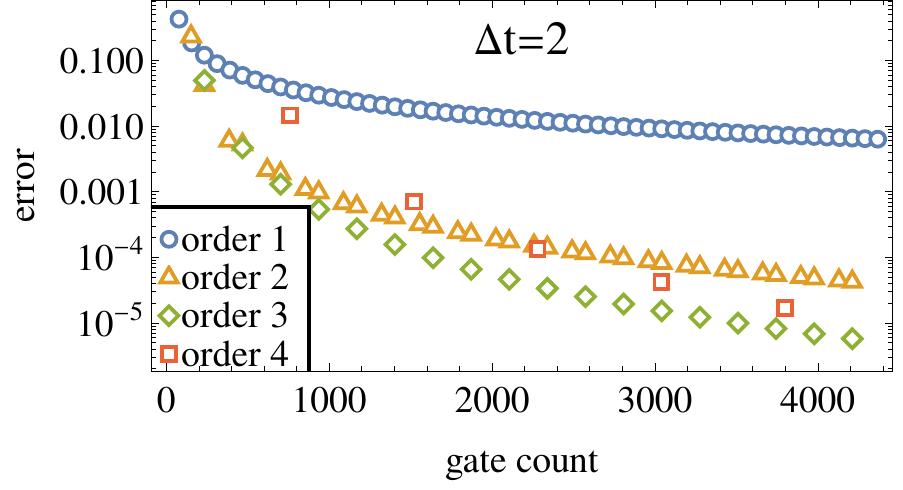}
    \caption{Naive Trotterisation on an $\text{H}_2$ molecule's propagator for $\Delta t=2$
        up to the fourth order. 
        Each order show Trotterisations with various trotter number ($n$), consistently increasing from
        $n=1$ up to $n=56$ for the first order; thus $n$ coincides with the number of markers.
        The error is defined as the matrix distance (\Cref{eq:distance}) between
        the propagator and the Trotterised circuit.  We refer
        to~\Cref{sec:trotter} for details on the Trotter formula, the circuits, and circuit
        simplification.
}
    \label{fig:gatecount}
\end{figure}

The Suzuki-Trotter formula approximates $U(\Delta t)$ by the products of the
exponents, in which the error term is improved by slicing the total time
$\Delta t$ into $n$ uniform subintervals, also known as \emph{Trotter number}.
Increasing $n$ results in a larger gate count, trading off with an improved
error (\Cref{eq:distance}), as demonstrated in \Cref{fig:gatecount}.  It is
worth mentioning that, the gate count can be improved with more rigorous gate
simplification by taking into account the ordering, which we do not consider
here.  Nevertheless, by using naive Trotterisation and simple gate
removals/simplifications, from \Cref{fig:gatecount}, we need around $900$ gates
to obtain an error of $0.001$, which is quite immense for a 4 qubit problem!

In this paper, we demonstate our circuit synthesis techniques by creating compact circuits for
$\text H_2$ propagators; we do so for various durations $\Delta t$. We take advantage of
the block diagonal structure of propagators to synthesise very shallow circuits
($\leq$18 gates) constrained to a subspace, with a target error of less than
0.001 for various $\Delta t$. Our sister paper~\cite{sister} shows that circuit
synthesis in a subspace is more efficient than in the entire space, given that
the unitary has a block diagonal form. In the following, we summarise the idea
of the subspace compilation technique; we refer to our sister
paper~\cite{sister} for further details.

\begin{figure}[h]
\begin{quantikz}
    \lstick{$\ket0^{\otimes d}_R$} & \gate[2]{P}\qwbundle{} & \gate{U(\Delta t)} & \gate[style={fill=black!10}]{\mathcal C(\theta)} & \gate[2]{P^\dagger} & \meter{}\qwbundle{} \\
\lstick{$\ket0^{\otimes m}_A$} &   \qwbundle{}&\qw &  \qw  &  &\meter{}\qwbundle{}
\end{quantikz}\caption{The circuit setup for computing the cost function.
    A maximally entangled state $\ket\Phi_{RA}$ (\Cref{eq:subspace_phi}) is
    prepared in registers $R$-$A$ by applying a preparation operator $P$ from
    a zero state $\ket{0}^{\otimes(d+m)}$. This technique is a generalisation
    of the circuit compilation introduced in \cite{sharma2020noise}.
    The propagator $U(\Delta t)$ is applied to the first register; it is the
    unitary target, \eg a long canonical circuit obtained by Trotterisation.
    Then, a parameterised ansatz $\mathcal C(\vec\theta)$ is applied, followed by
    measurement in basis $\ket{\Phi}_{RA}$. The measurement outcomes are
    accounted for in the cost function.} \label{fig:subspace_setup}
\end{figure}
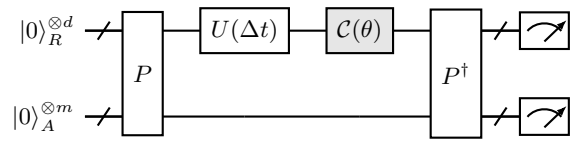

Given a propagator $U(\Delta t)$, we synthesise a hardware-efficient circuit
that approximates it by means of VQA.  Ideally, the operator $U(\Delta t)$ can
be implemented on the quantum computer with a great accuracy. For instance,
$U(\Delta t)$ is a high order Trotterisation with a very deep circuit.  Our VQA
is aimed at discovering a much shorter circuit that approximates $U(\Delta t)$
in the full Hilbert space and in a subspace.  The key to our circuit synthesis is evaluation of
the cost function shown in \Cref{fig:subspace_setup}.

Let $\mathcal H_S$ be the subspace of $\mathcal H$ that preserves the relevant occupation number; this
method requires $m$ ancillae, where $m=\lceil\log_2(\dim \mathcal H_S)\rceil$.
By referring to \Cref{fig:subspace_setup}, the core of our circuit synthesis technique is 
the minimization of  
\begin{equation}\label{eq:cost_compile}
    \min_{\vec\theta} (1-\bra*{\Phi}_{RA} U(\Delta t) \mathcal C(\vec\theta) \ket{\Phi}_{RA}),
\end{equation}
where $\ket{\Phi}_{RA}$ is maximally entangled state between spaces $R$ and $A$, 
namely,
\begin{equation}\label{eq:subspace_phi}
    \ket{\Phi}_{RA}=\frac{1}{\sqrt{2^m}}\sum_{s_j\in S}\ket{s_j}_R\ket{j}_A.
\end{equation}
It means $\Tr_A{(\ketbra\Phi)}_{RA}=I_R$ and $\Tr_R{(\ketbra\Phi)}_{RA}=I_A$, where $I$
is an identity matrix. 
Here, the basis $\{\ket{s_j}_R\}$ spans $\mathcal H_S$, which is the subspace of $\mathcal H$
spanned by basis $\{\ket{s_j j}_{RA}\}$.
Upon convergence of the cost function (\Cref{eq:cost_compile}) over a set of 
parameters $\vec\theta$, the time evolution 
operator is approximated by $U(\Delta t)\approx^{\epsilon}\mathcal C(\vec\theta)^\dagger$
with some error $\epsilon$. We define the error $\epsilon$ as a matrix distance between $U(\Delta t)$
and $\mathcal C(\vec\theta)^\dagger$; the definition of $\epsilon$
is elaborated later in \Cref{sec:res_the_time-evolution}.

\section{Discussion on the choice of optimisation}
\label{sec:discussion_on}

The performance of a VQA scheme crucially depends on the setup of the
optimiser. We outline the core aspects of our optimization techniques
below, which are elaborated in greater detail in our sister
paper~\cite{sister}, to which we refer the interested reader.

$\emph{Gate pool}$. We use a large set of gates comprising parameterised
single-qubit and controlled rotations assuming full connectivity, namely
$\{R\sigma_i(\theta),C_i(R\sigma_j(\theta))\}_{\sigma\in\{x,y,z\}}$, where
$i,j\in\{0,\cdots,n-1\}$ and $i\neq j$, for a $n$-qubit problem. The parameter sets are dense,
$\theta\in\mathbb R$, since most experiments allow continuous parameters.
Notice that the number of parameters coincides with the number of gates. 


\emph{Adaptive ansatz}.  The ansatz is constructed by trying out several
candidates of random ans\"atze for multiple rounds. On each round, every ansatz
candidate is initialized by a random circuit, comprising legitimate random
gates, which are in turn initialized with random parameters to be optimised.
The best candidate is chosen by the lowest cost function, which then continues
to evolve through the optimisation. This technique is inspired by the ADAPT
technique~\cite{grimsley2019adaptive}, in which the global minimum is approached
with small steps along the evolving ansatz.  For the ground state minimization,
the initial state is a product state with the correct occupation number; our
choice is $\ket{\psi_0}=\ket{0\cdots001\cdots11}$, by least significant bit
convention, where the number of ones coincides with the occupation number.  We
refer to the \emph{random search} technique in~\cite{sister} for further
details.


\emph{Removal of superfluous gates}.  At each step of ansatz evolution,
judicious deletions are performed on the ansatz before evolving further. This
step is crucial, since superfluous gates can trap the system in the local minima,
and a large number of optimisation parameters can induce a Barren plateau.  Our
sister paper~\cite{sister} elaborates on this removal technique in Section
III.D.

\emph{Parameter optimisation}.  We employ the natural gradient descent (or
imaginary-time evolution)
method~\cite{mcardle2019variational} to train
our parameters. The free parameters $\vec\theta$ are updated as follows
\begin{equation}\label{eq:training_param}
\vec\theta'= \vec\theta - \lambda F(\vec\theta)^{-1} \nabla C(\vec\theta),
\end{equation}
where $\lambda$ is the gradient step, $\nabla C(\vec\theta)$ is
gradient of the cost, and  
\begin{equation}\label{eq:fisher} \begin{split}
    F_{ij}=&4\Re\left[\braket{\partial_i\psi(\vec\theta)}{\partial_j\psi(\vec\theta)}
\right. \\ &\left.  -\braket{\partial_i\psi(\vec\theta)}{\psi(\vec\theta)}
\braket{\psi(\vec\theta)}{\partial_i\psi(\vec\theta)}\right] \end{split}
\end{equation} 
is the \emph{quantum Fisher information metric}~\cite{stokes2020quantum,meyer2021fisher},
also known as \emph{quantum metric tensor}.
Note that the equation above 
holds only for pure states.  The metric tensor gives additional information
in the state space from perturbation in the parameter space; this improves 
the optimisation performance as demonstrated
in~\cite{mcardle2019variational}.  Also, to speed up optimisation, 
we exponentially increase lambda within a single iteration until we arrive at a
local minimum along the natural gradient direction $F^{-1}(\vec\theta)\nabla C(\vec\theta)$.

\emph{Hyperparameters}. Several hyperparameters are set according to the
problem, usually by the number of qubits and estimated difficulty.  These include the number of
random gates inserted at once, the proposition of one- and two-qubit gates insertion,
convergence criteria, the gradient step ($\lambda$ in \Cref{eq:fisher}), the regularization parameter for $F^{-1}(\vec\theta)$
(\Cref{eq:training_param}), and some parameters relevant to the gates removal
subroutines. Moreover, some hyperparameters are updated when the optimisation fails
to lower the cost function.

\section{Results and analysis}\label{sec:results_and}

The results reported here are accessible in our GitHub
repository~\cite{cicacica}.  It is important to note that we do not consider
noise in our simulations.  We use \texttt{QuESTLink}~\cite{jones2020questlink}
to emulate our algorithms, which is the \texttt{Mathematica} interface to a compiled form of the
\texttt{QuEST}~\cite{jones2019quest} toolset (a highly efficient c codebase). We employ a GPU to accelerate our
computations. As \texttt{Python} is popular in the relevant communities, it is
worth mentioning that one can also reproduce our simulations using
\texttt{pyQuEST}~\cite{pyquest}, which is the \texttt{Python} interface to
\texttt{QuEST}. 

\subsection{Ground state estimations}
\label{sec:res_ground_state}

\begin{figure}[t]
\centering
\includegraphics[width=.85\columnwidth]{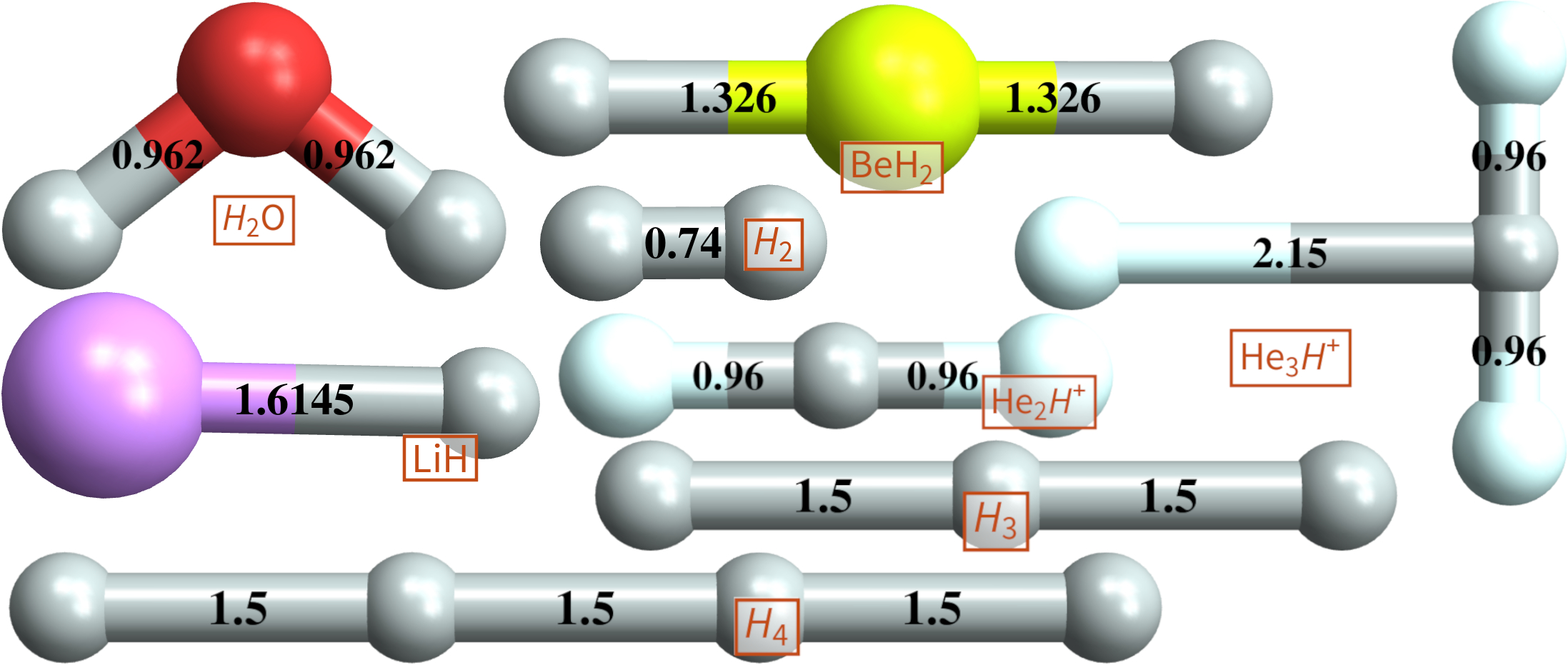}
\caption{
\label{fig:molecules}
Geometry of each molecule used in this work.
The units are displayed in Angstrom (\AA).
}
\end{figure}

We test our optimisation technique to estimate the ground states of small
molecules that can be represented by up to 14 qubits. We use the stable
geometry for the molecules and constant distance for the (artificial) hydrogen
chains. The geometry of each molecule is shown in \Cref{fig:molecules}.

\begin{figure}[t]
    \centering
    \subfloat[\label{fig:vqe1}]{
        \resizebox{4.8cm}{!}{\includegraphics{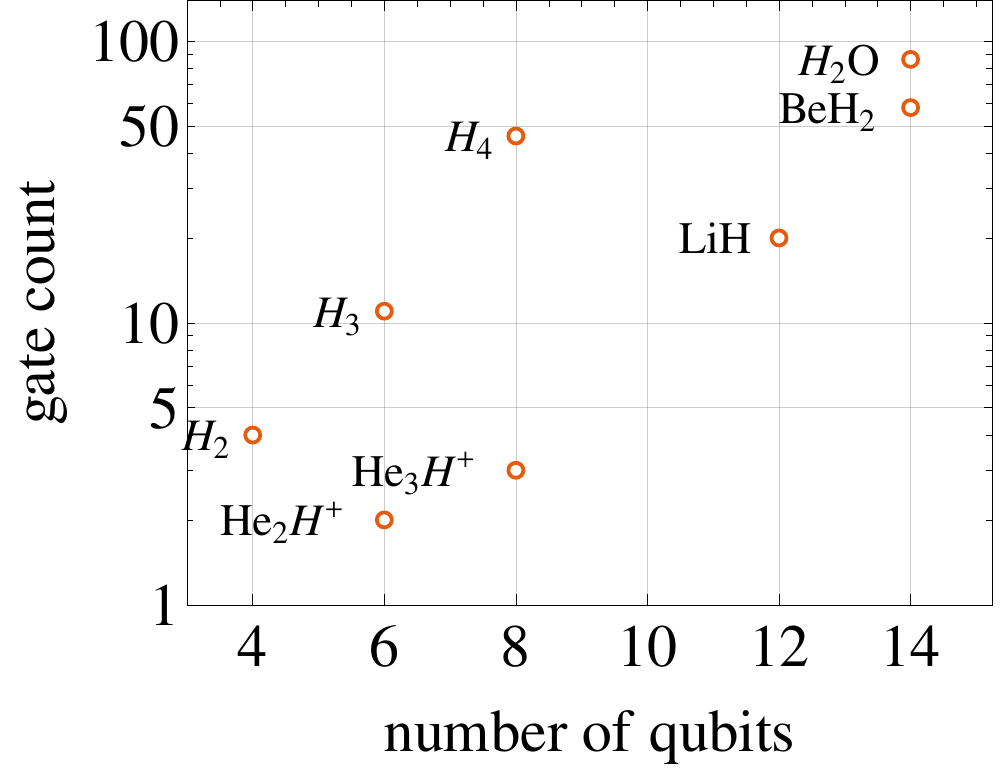}}}
        \subfloat[\label{fig:vqe2}]{
    \adjustbox{valign=b}{
        \resizebox{3.8cm}{!}{
    \begin{tabular}{|M{1cm}|M{.8cm}|M{.8cm}|M{1cm}|}
    \hline
    Mole-cule & Pauli terms &Two-qubit gate & Single-qubit gate \\ \hline
    $\text H_2$ & 15 & 3& 1\\ \hline
    $\text H_3$ & 62& 8& 3  \\ \hline
    $\text H_4$ & 185 &36& 10 \\ \hline
    $\text{He}_2\text H^+$ & 62& 1 & 1 \\ \hline
    $\text{He}_3\text H^+$ & 193 & 1& 2 \\ \hline
    LiH & 631 & 12 & 8 \\ \hline
    $\text{BeH}_2$ & 666 & 39 & 19   \\ \hline
    $\text{H}_2\text O$ & 1086 &49 & 37  \\ 
    \hline
  \end{tabular}}
}
}
\caption{
\label{fig:vqe_main}
    The gate count obtained by our optimiser to prepare the ground state within chemical accuracy.
    The number of parameters coincides with the gate count.
    (a) The gate count plot versus the number of qubits. (b) The gate count for two- and one-qubit gates compared
    to the number of Pauli terms in the Hamiltonian. 
}
\end{figure}

We terminate the optimization when the updated energy fails to lower the cost
function by more than $\delta=10^{-6}$.  Each molecule is run for ten trials. Our
best results --- by the total gate count --- among the trials, within chemical
accuracy, are shown in \Cref{fig:vqe_main}.  


\begin{table}[h]
    \resizebox{\columnwidth}{!}{
        \bgroup
\def\arraystretch{1.1}%
\begin{tabular}{|C{.9cm}|C{1cm}|C{.9cm}|C{7.5cm}|}
    \hline
    Mole-cule & Gate count & Param count & Description  \\ \hline
    \multirow{2}{*}{$\text H_2$} & 125 &2 & ADAPT-VQE~\cite{claudino2020benchmarking}. \\ \cline{2-4}
    & 105 &70 & Fixed ansatz~\cite{tkachenko2021correlation}, with 35 CNOT and 70 $z$-rotation gates, has 8 qubits, with STO-6G basis, and nearest-neighbour connectivity. \\ \hline
    \multirow{3}{*}{$\text H_4$} & $>150$ &26 & AVQITE~\cite{gomes2021adaptive}, based CNOT count with cyclic nearest-neighbour connectivity, and tapered to 6 qubits. \\ \cline{2-4}
    & 279 &248 & MoG-VQE~\cite{chivilikhin2020mog}, with bond distance 1.2\AA, 31 CNOT and 248 single rotations gates. \\ \cline{2-4}
    & $>$2208 &11 & qubit-ADAPT~\cite{tang2021qubit}, based on CNOT count, and within accuracy $10^{-14}$. \\ \hline
    \multirow{3}{*}{LiH} & 108 &72 & Fixed ansatz\cite{tkachenko2021correlation}, with 36 CNOT and 72 $z$-rotation gates, tapered to 10 qubits, and has nearest-neighbour connectivity.\\ \cline{2-4}
    & 108 &96 & MoG-VQE~\cite{chivilikhin2020mog}, with 12 CNOT and 96 single rotation gates.\\ \cline{2-4}
    & $>6824$ &30 & qubit-ADAPT\cite{tang2021qubit}, based on CNOT count, and within accuracy $10^{-10}$.\\ \hline
    \multirow{3}{*}{$\text{BeH}_2$} & $>$400 &46 & AVQITE~\cite{gomes2021adaptive}, based on CNOT count with cyclic nearest-neighbour connectivity, and tapered to 12 qubits.\\ \cline{2-4}
    & 81 &72 & MoG-VQE~\cite{chivilikhin2020mog}, with 9 CNOT and 72 single rotation gates, and tapered to 8 qubits. \\ \hline
    \multirow{2}{*}{$\text{H}_2\text O$} & $>1550$ &84 & 4-UpCCGSD-PECT ansatz~\cite{sim2021adaptive}, and count based on circuit depth. \\ \cline{2-4} 
    & $>$100 &5 & AVQITE~\cite{gomes2021adaptive}, based on CNOT count with cyclic neares-neighbour connectivity, and tapered to 8 qubits. \\ 
    \hline
  \end{tabular}
  \egroup
  }
  \caption{\label{tab:vqe_others}The gate and parameter counts from various methods. 
 Methodological differences to ours are summarised in the description column. We select these results because of similarity in the molecular geometry and data availability.}
\end{table}

\begin{figure}[!h]
    \resizebox{\columnwidth}{!}{
    \input{lih}}
    \caption{\label{fig:explicit} The LiH ground state preparation circuit synthesised by
        our algorithm. The LiH molecule is commonly used to benchmark VQE algorithms. 
        The circuit is rounded to the multiple of $\pi$ or to two-decimal places; it has energy error of $\varepsilon=1.4\times 10^{-3}$ Hartree. 
        Three qubits remain untouched; two of them could have been manually identified as redundant by tapering them off~\cite{bravyi2017tapering}, \ie reducing the number of qubits by preserving some symmetries in the Hamiltonian.
    }
\end{figure}
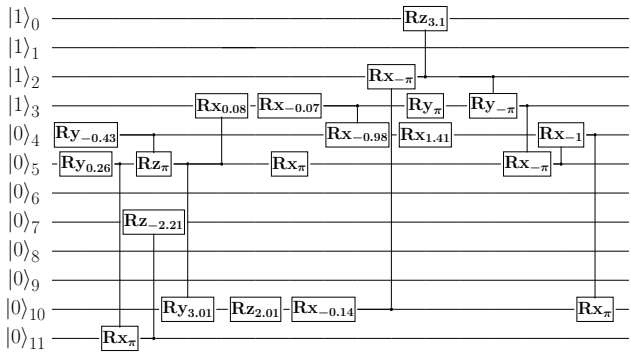


The resulting gate counts are remarkably low for some small molecules, such
as $\text H_2$, $\text{He}_2\text H^+$, and $\text{He}_3\text H^+$. The initial
states are already somewhat close to the ground state in these cases. We note that, for the
$\text{He}_2\text H^+$ case, the initial state can easily be transformed into the ground state:
we start with state $\ket{011111}$, where the ground state is
$0.071669\ket{011111}+0.997428\ket{111101}$. That is achievable with a hadamard and 
a controlled-rotation gates. Our optimiser 
discovered circuit $Rx_5(-0.1453)C_5[Rx_1(\pi)]$ to approximate the ground 
state of $\text{He}_2\text H^+$ with an energy error magnitude of $10^{-10}$.

To put our results into context, we compare them with the results of various
methods tabulated in \Cref{tab:vqe_others}. We emphasise that this comparison
is not straightforward due to a vast landscape of conventions. We observe
differences in gate pools, unreported total gate count, objective functions,
accuracies, molecular geometries, and Hamiltonian mappings may lead to unfair
comparisons.  For instance, the results from MoG-VQE~\cite{chivilikhin2020mog}
show small numbers of CNOT gates but large numbers of single-qubit gates. This is no doubt
because the MoG-VQE method focuses on reducing CNOT gates while single-qubit
gates are regarded as plentiful and placed without simplification.  Nonetheless, in most
cases, we find lower numbers of total gates compared to other results even
when we decompose our circuits into CNOT and single rotation gates.  For
instance, \Cref{fig:explicit} shows our circuit to prepare the ground state of
LiH, which comprises only 20 gates.

\begin{figure}[t]
    \includegraphics[scale=0.6]{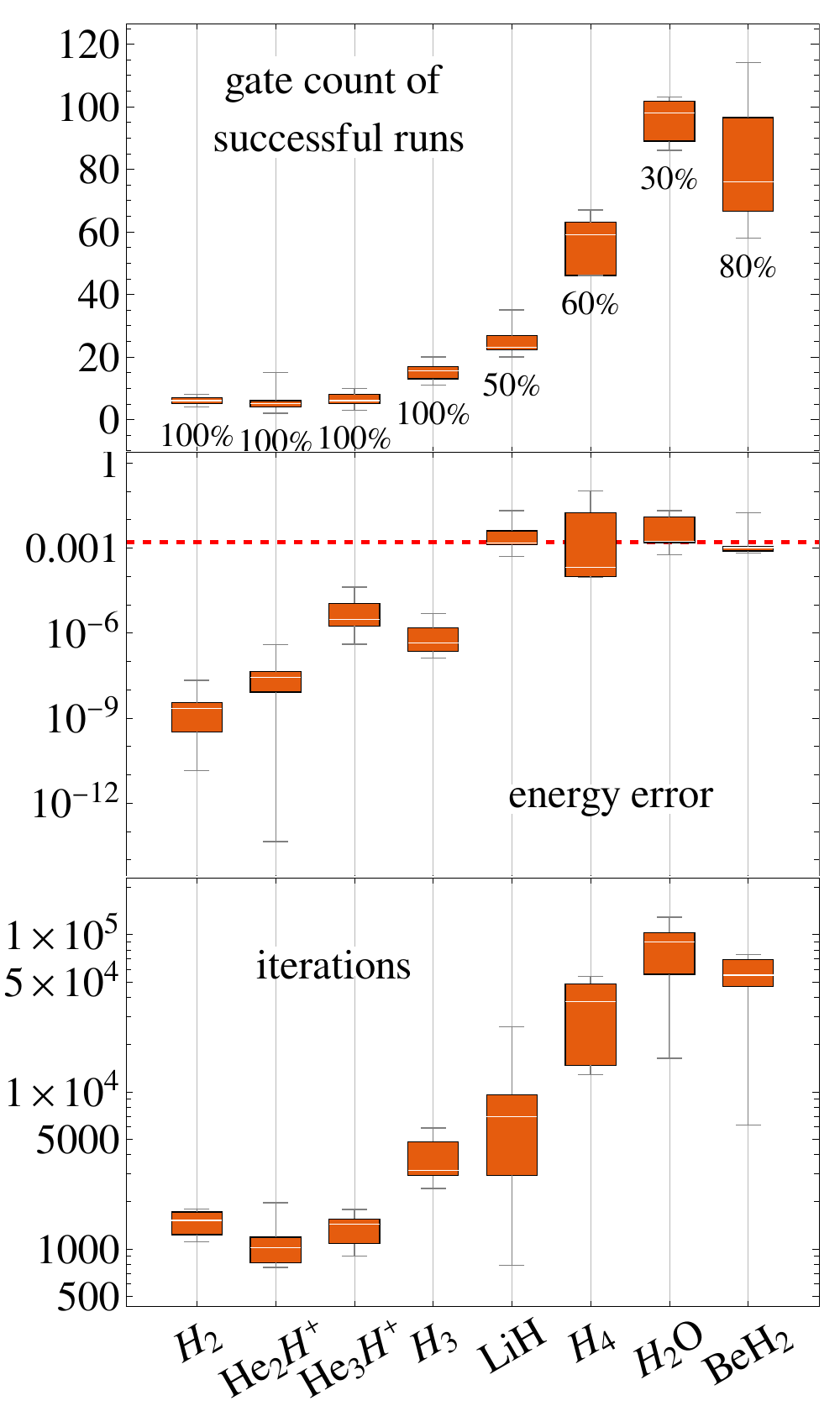}
    \caption{\label{fig:whiskers} 
Optimisation results from all trials for each molecule. The \emph{gate
count} section shows the successful run defined by energy errors within chemical
accuracy. The \emph{success rate}, \ie percentage of trials that converge below
the threshold $\delta=10^{-6}$, are shown below each whisker. The water
molecule shows the lowest rate of successful runs. In the \emph{energy
error} section, the dashed red line indicates chemical accuracy. The large
molecules show more results converged to the local minima. Finally, the
\emph{iterations} section shows the total gradient evaluations, including
trying all candidates and judicious deletions.  }
\end{figure}

Although our results show small gate counts, our ansatz choice comes with a
drawback due to the growing number of parameters. For instance, in constructing
the ansatz, sometimes, the appropriate gates are discovered by creating a large
random circuit followed by judicious deletions. Since large circuits come with
many parameters, our method might be prone to a barren plateau on larger
scales. On the other hand, the techniques that employ coupled-cluster family
operators such as qubit-ADAPT and ADAPT-VQE show large gate counts but have a
small number of parameters; these techniques might be more scalable than ours.

In the following, we discuss the factors influencing the difficulty of
preparing the ground state and how they affect the performance of our
optimisations.  First, we know that the difficulty in discovering the ground
state increases with the size of the molecule and the number of terms in the
Hamiltonians, but, they are not the only factors.  For instance, the ground
state of $\text{He}_3\text H^+$, which is represented using eight qubits,
requires only three gates; in contrast to $\text H_3$, which has six qubits but
requires 11 gates.  Also, for an equally big 8-qubit problem, $\text H_4$
needs 46 gates accompanied with a lower success rate and needs more
iterations --- see \Cref{fig:whiskers}, even though it has fewer Pauli terms.
Since the task of estimating the ground state problem is proven to be 
QMA-complete~\cite{kempe2004the}, rigourously quantifying the optimisation difficulty of
a Hamiltonian of practical scale may remain an unfeasibly hard challenge.

Intuitively, if we knew that the ground state was highly correlated, we would
expect the optimisation to be difficult. A highly correlated state usually
corresponds to a highly entangled state, which requires a sophisticated circuit
with numerous entangling gates to produce.  Thus, circuits preparing highly
entangled ground states are more difficult to synthesise as they need deep
circuits and (or) large number of parameters which are likely to yield flat
gradient manifolds.  Therefore, in the following, we quantify the entanglement
of the target state and evaluate its impact on the performance of our
optimiser.

Having the advantage of an even number of qubits in our molecules, we can
straightforwardly use the bipartite entanglement measure between two chosen
subsystems.  A simple yet powerful measure for pure states is the 
\emph{Schmidt measure (Hartley strength)}\cite{jaeger2007quantum}: 
\begin{equation}
    E(\ket*\psi)=\log_2\left(\rank(\ket*{\psi_{sch}})\right), 
    \label{eq:hartley}
\end{equation}
where $\ket*{\psi_{sch}}$ is the Schmidt form of $\ket*\psi$. 
The unit of Hartley strength is known as ``ebit'', where Bell states correspond
to one ebit. 

\begin{figure}[h]
    \centering
    \includegraphics[scale=.65]{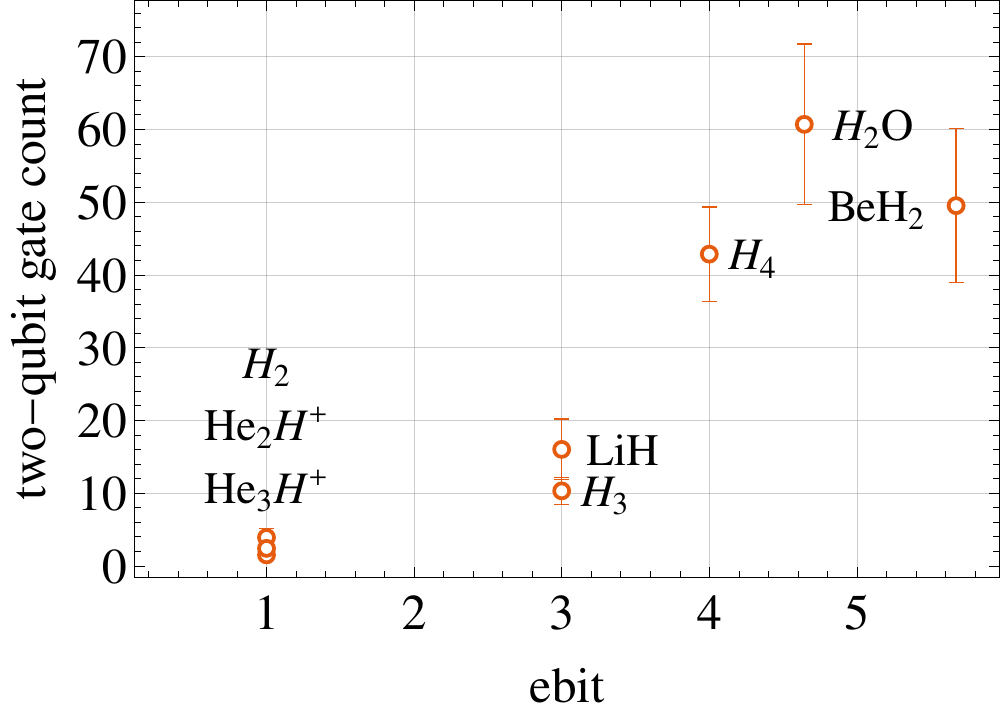}
    \caption{Schmidt measure of the ground state of each molecule (in ebit)
        plotted against the average two-qubit gate count of successful runs. It shows that
        states with high entanglement need more entangling gates to prepare. 
    }
    \label{fig:hartley}
\end{figure}

Indeed, as shown \Cref{fig:hartley}, the entanglement measure shows a stronger
correlation to the entangling gate count than the number of qubits or the Pauli
terms.  For instance, the two-qubit gate counts of the molecules $\text H_2,
\text{He}_2\text H^+$, and $\text{He}_3\text H^+$ are almost identical for the
same Hartley strength. Consistent results also shown for $\text H_3$ and LiH,
although LiH was more difficult for us to optimise, with only a 50\% success
rate (the error within chemical accuracy). In addition, $\text H_4$ is expected
to be more difficult than LiH and $\text H_3$, which is also reflected in the
optimisation performance. However, in the case of $\text H_2\text O$ and
$\text{BeH}_2$, our optimiser found that $\text H_2\text O$ is harder than
$\text{BeH}_2$.  The performance of our optimiser relies on hyperparameters
which are manually set.  It might be that the hyperparameters we chose for $\text
H_2\text O$ are not optimal for this case. 

Our optimiser has successfully learned shallow circuits that prepare the ground
states with gate counts much lower than $\Theta(n^3 4^n)$, which is a bound on
the number of two-qubit gates necessary to prepare an arbitrary $n$-dimensional
unitary gates~\cite{barenco1995elementary}. Moreover, some of our results
outperform results by other methods as shown in \Cref{tab:vqe_others}. However,
the performance of our optimiser relies on randomness and a good choice of
hyperparameters. For instance, in some cases, we insert a hundred random gates
in which 99 of them are deleted and only one remains; this is a significant
overhead in guessing a single gate.  Many trials are sometimes required to
guess the proper hyperparameters as well.  Therefore, we do anticipate the need for adaptions and refinements in scaling the method to much larger molecules. 


\subsection{The time evolution operator circuits in the subspace and full space}\label{sec:res_the_time-evolution}

Using the same optimisation technique, we synthesise circuits that implement time
evolution operators of the $\text H_2$ molecule with various time steps $\Delta
t$.  The optimisations are performed in the entire Hilbert space and in the
subspace that preserves the correct number of particles. Interestingly, as we
will see later, the resulting gate counts remain relatively constant while
increasing $\Delta t$.

First, we formalise the operator that quantifies the quality of the synthesised circuits. 
It is natural to compare two matrices by a distance norm to quantify accuracy. By referring to
\Cref{fig:subspace_setup}, the error of an approximation $U(\Delta
t)\approx^{\epsilon}C^\dagger (\vec\theta)$ that is obtained by recompilation in 
a subspace $\mathcal S\subseteq \mathcal H$ is defined as
\begin{equation}\label{eq:distance}
    \epsilon=\min_\phi \norm{\Pi_{\mathcal S}\left(U(\Delta t)
    -e^{i\phi}\mathcal C(\vec\theta)^\dagger\right)\Pi_{\mathcal S}}_\infty,
\end{equation}
where $\Pi_{\mathcal S}$ is the projector to the subspace $\mathcal S$, and
$\norm{.}_\infty$ is Schatten-$\infty$ norm operator defined as 
\begin{equation}
    \norm{A}_\infty=\max \{ \norm{A v}: v\in \mathcal S \}, 
\end{equation}
which can be obtained by taking the largest magnitude eigenvalue.
For practicality, we say ``subspace error'', $\epsilon_{\mathcal S}$, if defined over $\mathcal S$ 
and ``full space error'', $\epsilon_{\mathcal H}$, if defined over $\mathcal H$. 
In the context of circuit synthesis, we interchangeably use error and distance for clarity.

\begin{figure}[ht]
    \subfloat[\label{fig:subspace_distance}]{
    \centering
    \resizebox{.8\columnwidth}{!}{
    \includegraphics{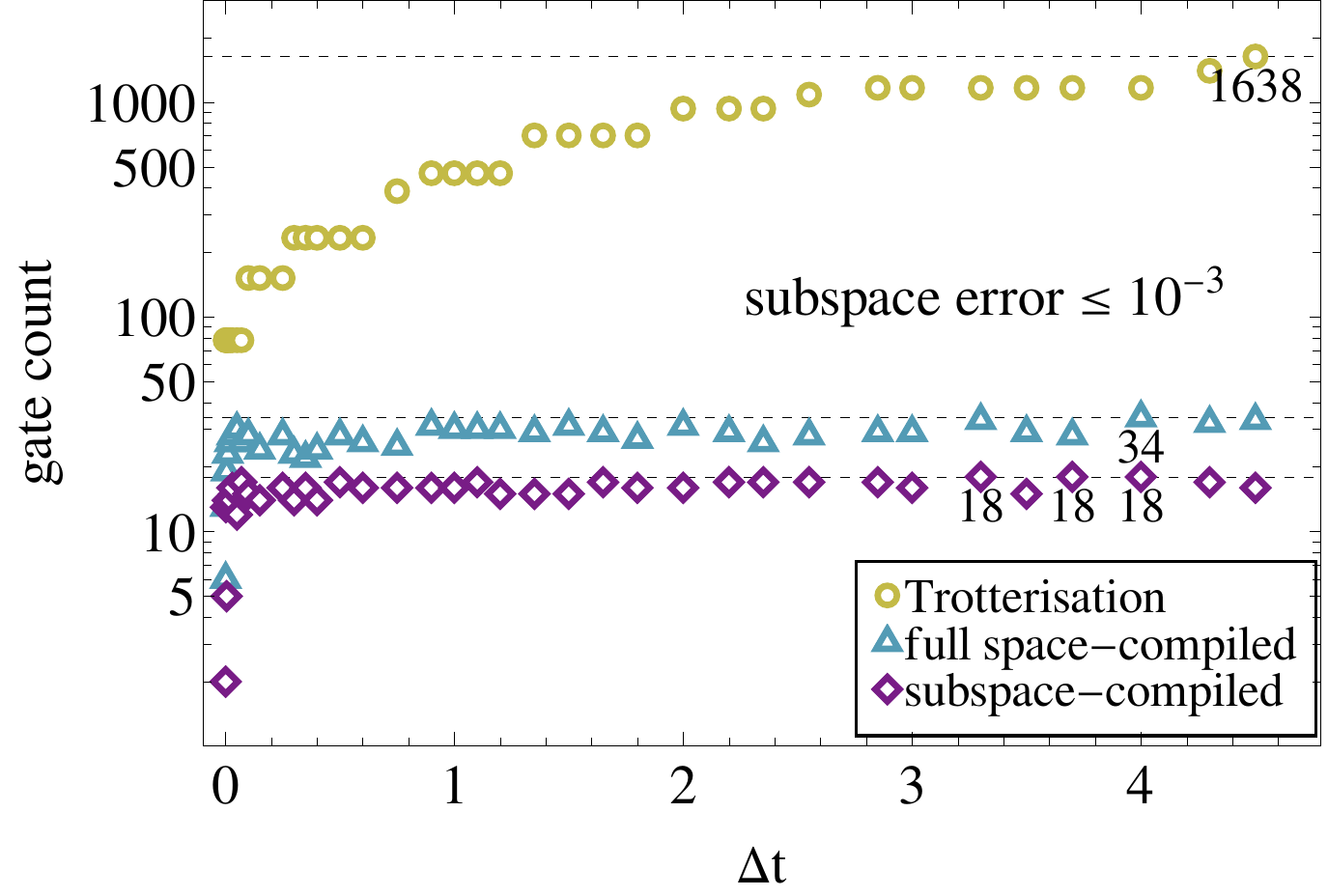}}}
    \hspace{1.5cm}
    \subfloat[\label{fig:fullspace_distance}]{
    \centering
    \resizebox{.8\columnwidth}{!}{
    \includegraphics{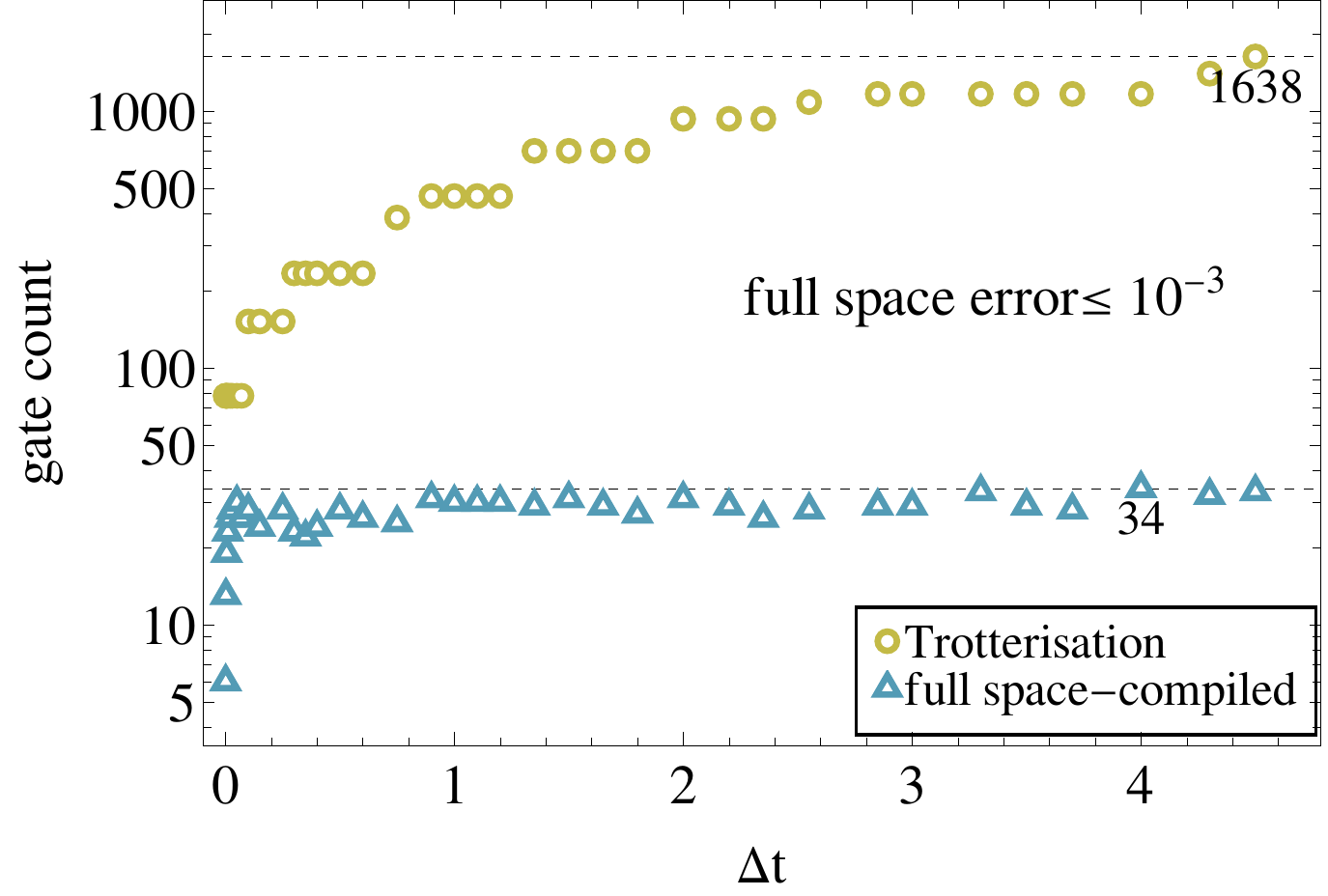}}}
    \caption{
        \label{fig:dyn_main}
        Semilogarithmic plots of the gate counts implementing circuits approximating $U(\Delta t)$
        with (a) subspace error $\epsilon_{\mathcal S}\leq
        10^{-3}$ and (b) full space error $\epsilon_{\mathcal H}\leq 10^{-3}$.
        The compilation are done in the entire Hilbert space (full space-compiled) and 
        in the subspace with occupation number two.
        The Trotterisation gate counts are obtained by taking the
        minimum count over various Trotterisations with different trotter
        numbers and orders as done in \Cref{fig:gatecount}; the total
        Trotterisations considered to obtain this plot are shown in
        \Cref{fig:all_trotter}, in \Cref{sec:trotter}.  For the compilation
        methods, we select the least count among the ten trials. The dashed
        lines indicate the largest count on each method, where the labeled
        markers show cases with the largest count.
    }
\end{figure}

We perform optimisations in the subspace $\mathcal S$ and in the full space
$\mathcal H$, where $\mathcal S$ is the subspace with occupation number two ---
as present in $\text H_2$ molecule; thus, $\dim(\mathcal S)=6$ and
$\dim(\mathcal H)=16$.  We set the cost convergence condition to $\delta\leq
10^{-7}$, where the cost is defined in \Cref{eq:cost_compile}. For such a
$\delta$, we obtain a distance in of the order $10^{-4}$. We synthesise
propagator circuits for 37 different $\Delta t$ values, ranging from 0.001 to
4.5. For each $\Delta t$, we try ten independent calculations. The best results
of all trials with errors $\epsilon_{\mathcal H},\epsilon_{\mathcal S}\leq
10^{-3}$ are shown in \Cref{fig:dyn_main}. In that figure, we also compare
our results with the Trotterisation method.

Interestingly, \Cref{fig:dyn_main} shows that
the circuits synthesised by our optimiser have a relatively constant number of
gates, for both the compilation in the subspace and the full space. Moreover,
the gate counts we have obtained are very low: no more than 18 for the subspace
compilations and no more than 34 for the full space compilations; those are
significantly smaller than the gate counts we have obtained via Trotterisation,
which grow exponentially with $\Delta t$.

\Cref{fig:subspace_distance} shows that subspace-compiled gate counts are
always smaller than the full space-compiled ones. This is consistent with an
observation in our sister paper \cite{sister}: the subspace compilation is more
efficient and performs better than full space compilation. One can see it as
reducing the problem size by compiling in the subspace. We provide details on
the performance of our circuit synthesis in \Cref{sec:recompilation}.

In this numerical experiment, we observe that the complexity of the unitary is
practically the same for different $\Delta t$ with our technique. This may be
due to the nature of parameterised ansatz optimisation that trains the
parameters within a particular unitary structure. For instance, this technique
was tested on a dense random unitary in our sister paper, resulting in significantly
more complex optimisations and showing much larger gate counts. Fortunately,
the time-evolution operators have a constant unitary structure for any $\Delta
t$.  Its sparse structure and block diagonal form are ideal to our compilation
techniques. 

Our results may provide a significant advantage in some practical applications
by the following schemes. First, a long-duration time-evolution simulation by
repeated application of a propagator is more feasible due to the shallow
circuit of each iterate. Second, one can compile a long-duration
time-evolution propagator by recursive compilation to obtain a short circuit to
simulate a longer duration. The advantage is even more significant for
reactions that preserve the number of particles. Finally, it possesses a
potential in first-quantised
simulations~\cite{fleck1976time,kassal2008polynomial,chan2022grid}. For
instance, \citet{chan2022grid} have successfully predicted oscillations in the
Helium dimer using the first-quantised method.  Their simulation extensively
employs propagators for each potential term, spatial grid, and time grid; thus,
maintaining shallow circuits of the propagators can significantly improve the
efficiency, which are expected to assist in the prediction of the behaviour of
larger molecules in the future.

\section{Conclusion and further discussion}
\label{sec:conclusion_and}

We implemented optimisation techniques studied in our sister
paper\cite{sister}: (1) ground state estimations of molecules $\text H_2$,
$\text H_3$, $\text H_4$, $\text{He}_2\text H^+$, $\text{He}_3\text H^+$, LiH,
$\text{BeH}_2$, and $\text{H}_2\text O$, and (2) synthesising circuits of the
$\text H_2$ molecule time-evolution operators for various time durations $\Delta
t$ in which optimisations are run in the subspace that preserves the number of
particles, as well as in the entire space. Key results from our
study here are the following. For problem (1), we obtained shallow circuits to
prepare the ground state of each molecule summarised in \Cref{fig:vqe_main}.
For problem (2), we synthesised shallow circuits with a practically constant
depth for all given $\Delta t$, which is summarised in \Cref{fig:dyn_main}. We found that
compilation in the subspace reduces the gate count versus
the full space compilations by a factor of about 2.

We can anticipate some of the challenges associated with scaling the technique to problems beyond the quantum advantage threshold. First, it might not
scale well since it relies on randomness to find gates that can lower the cost,
and the parameter count grows with the number of gates simultaneously inserted. This would likely necessitate a more nuanced protocol for gate introduction in place of the present purely-random heuristic. 
Second, the performance relies on the hyperparameters that are chosen manually. The solution, automatic selection of hyperparameters, would be an interesting machine learning task in its own right.
Regardless of these caveats, our optimiser has already demonstrated potential to solve small molecule chemistry
problems with remarkable efficiency.

As an additional note, it was challenging for us to comprehensively
characterize the outcomes of our method with the literature due to the lack of
a common framework and transparency.  For instance, in our opinion, the total
gate count is as important as the two-qubit gate count and the circuit depth
for the following reasons. First, by the state of the art of quantum computing,
\ie in the NISQ era, most experiments do not yet exploit the parallelism of
quantum gates. Second, performing CNOT gate and arbitrary controlled-$x$
rotations without fault-tolerance is equally difficult in experiments. Finally,
VQE is aimed at non-fault-tolerant devices such as NISQ devices. Hopefully,
the upcoming works in this area will show more transparent results and consider
the experimental situations to push the practical applications of NISQ forward.


\section{Acknowledgements}

We thank Daniel Marti Dafcik, Hans Hon Sang Chan, and B\'alint Koczor for the
discussions on the variational algorithms and chemistry problems.  We thank
Tyson Jones for developing \texttt{QuESTlink} and writing functionalities such
that we can conveniently implement our ideas.  C.G thanks Jonathan Conrad for
the proof reading. C.G and S.C.B acknowledge financial support from EPSRC Hub
grants under agreement No. EP/T001062/1, from the IARPA funded LogiQ
project, the EU flagship AQTION project, European Union’s Horizon 2020 research,
and innovation programme under grant agreement no.
951852 (QLSI).

\bibliography{main}

\pagebreak
\onecolumngrid
\appendix
\section{Trotterisation}\label{sec:trotter}

Given that the Hamiltonian can be represented as 
\[
    \sum_{k}c_k P_k,\quad\text{where }c_k\in\mathbb R,P_k\in\{X,Y,Z,I\}^{\otimes n}.
\]
We partition all of the Pauli terms $\{P_k\}$ into subsets $A$ and $B$ in which
elements are commuting to each other.  The following equations are the first
fourth orders of Suzuki-Trotter formulas\cite{trotter,seeley2012bravyi} used in this paper:
\begin{align}
    e^{(A+B)t}&\approx (e^{At/n}e^{Bt/n})^n+O(t^2)\\
    e^{(A+B)t}&\approx \underbrace{(e^{At/2n}e^{Bt/n}e^{At/2n})(e^{Bt/2n}e^{At/n}e^{Bt/2n})(e^{At/2n}e^{Bt/n}e^{At/2n})\ldots }_{n\,\text{terms}}   +O(t^3)\label{eq:t2}\\
e^{(A+B)t}&\approx (e^{\frac{7}{4}At/n}e^{\frac{2}{3}Bt/n}e^{\frac{3}{4}At/n}e^{\frac{-2}{3}Bt/n}e^{\frac{-1}{24}At/n}e^{Bt/n})^n+O(t^4)\\
e^{(A+B)t}&\approx \left(\prod_{i=1}^{5} e^{p_i At/2n} e^{p_i Bt/n}e^{p_i At/2n}\right)^n+O(t^5),\quad p_1=p_2=p_4=p_5=\frac{1}{4-4^{1/3}},p_3=1-4p_1. 
\end{align}
We use a slightly different ordering for the second order Trotter in \Cref{eq:t2} because
this ordering shows better error than the canonical one. 

We use the canonical exponentiating circuit form, for  
example,
\[
    e^{-i \alpha Y_0Z_1X_2Z_3}=
    \begin{quantikz}
        \lstick{$q_0$} & \gate{Rx(-\pi/2)} & \ctrl{1}& \qw & \qw & \qw &\qw&\qw & \ctrl{1}&\gate{Rx(\pi/2)}&\qw\\
        \lstick{$q_1$} & \qw              & \targ{} & \ctrl{1} &\qw&\qw&\qw & \ctrl{1}&\targ{}&\qw&\qw\\
        \lstick{$q_2$} & \gate{H}         & \qw     & \targ{} & \ctrl{1}&\qw&\ctrl{1}&\targ{}&\qw&\gate{H}&\qw \\
        \lstick{$q_3$} & \qw              & \qw     & \qw & \targ{}&\gate{Rz(2\alpha)}&\targ{}&\qw&\qw&\qw&\qw \\
    \end{quantikz}.
\]
The circuit is then simplified by the following operations:
\begin{gather*}
    CNOT_{i,j}CNOT_{i,j}=I, \\
    Rx_i(a) Rx_i(b) = Rx_i(a+b), \\
    Rx_i(+a) Rx_i(-a) = I, \\
    H_i H_i = I.
\end{gather*}
Note that we do not consider commutations in simplifying the gates, which can
improve the gate counts.

We generate Trotter circuits for various durations,
\begin{equation}\label{eq:deltat}
    \begin{split}
    \Delta t\in & \{
0.001, 0.002, 0.005, 0.007, 0.01, 0.02, 0.03, 0.035, 0.05, 0.07, 
0.1, 0.15, 0.25, 0.3, 0.35, 0.4, 0.5, 0.6,\\& 0.75, 0.9, 1, 1.1, 1.2, 
1.35, 1.5, 1.65, 1.8, 2, 2.2, 2.35, 2.55, 2.85, 3, 3.3, 3.5, 3.7, 4, 
4.3, 4.5\}.
\end{split}
\end{equation}

\begin{figure}[hbt]
    \includegraphics[scale=.45]{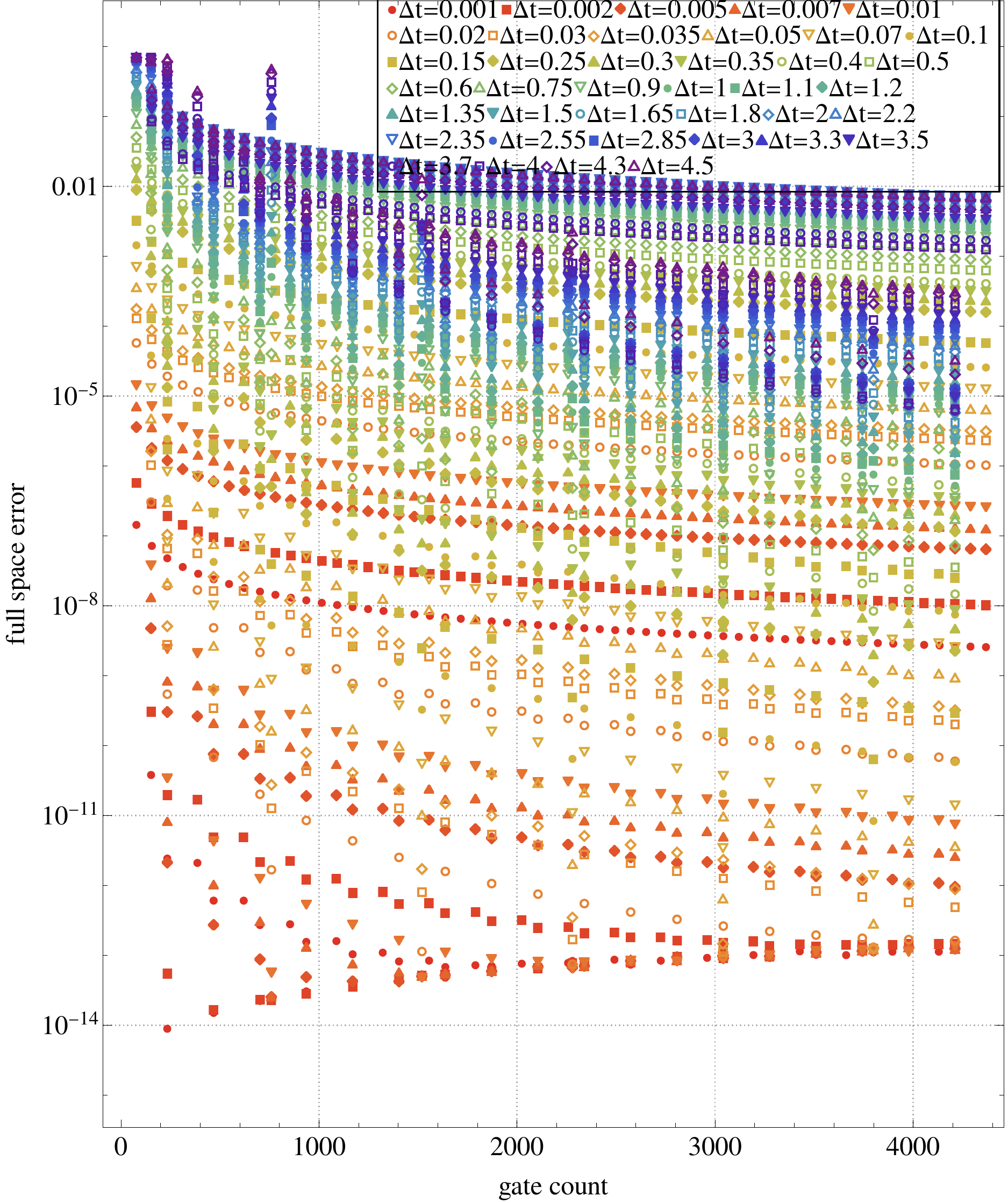}
    \includegraphics[scale=.45]{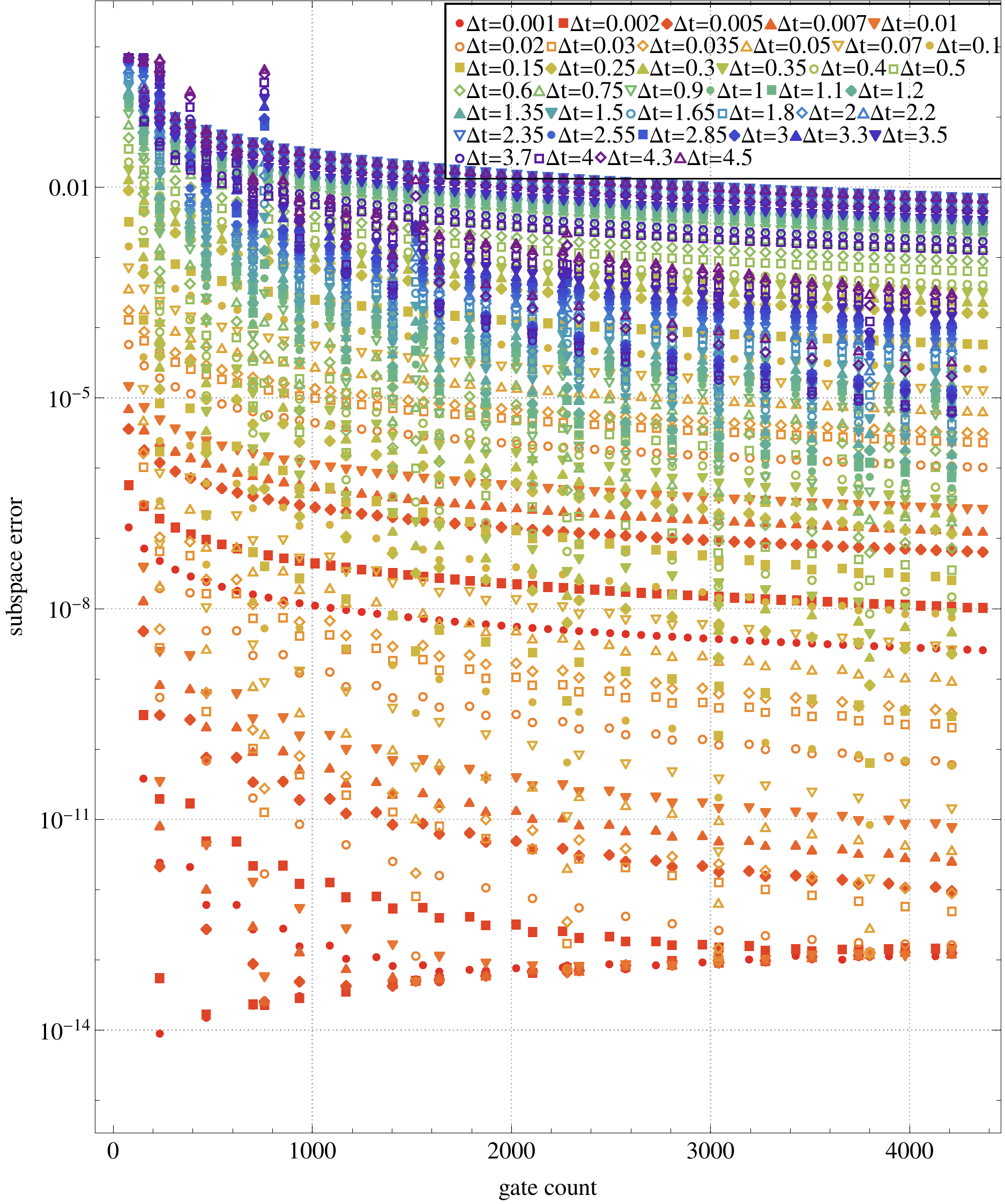}
    \caption{\label{fig:all_trotter}
        The full space and subspace errors of Trotter circuits by the gate counts.
        The data comprises various $\Delta t$ defined in \Cref{eq:deltat} and 
        various Trotter number $n$ defined in \Cref{eq:trotternum}.
    }
\end{figure}

For every $\Delta t$, we find Trotter circuits for the first fourth order with
various Trotter numbers $n$, where 
\begin{equation}\label{eq:trotternum}
\begin{split}
    n&\in\{1,\dots,56\}\quad \text{for the first order},\\
    n&\in\{1,\dots,36\}\quad \text{for the second order},\\
    n&\in\{1,\dots,18\}\quad \text{for the third order, and }\\
    n&\in\{1,\dots,5\}\quad \text{for the fourth order}.
\end{split}
\end{equation}
For each Trotter circuit, we compute the error for the full space and 
subspace as shown in \Cref{fig:all_trotter}. The error
is defined by the matrix distance to the exact matrix that is obtained by
\texttt{Mathematica} function \texttt{MatrixExp}. 

\section{Circuit synthesis of the time-evolution operator}\label{sec:recompilation}

\begin{figure}[h]
    \resizebox{.8\columnwidth}{!}{
    \includegraphics{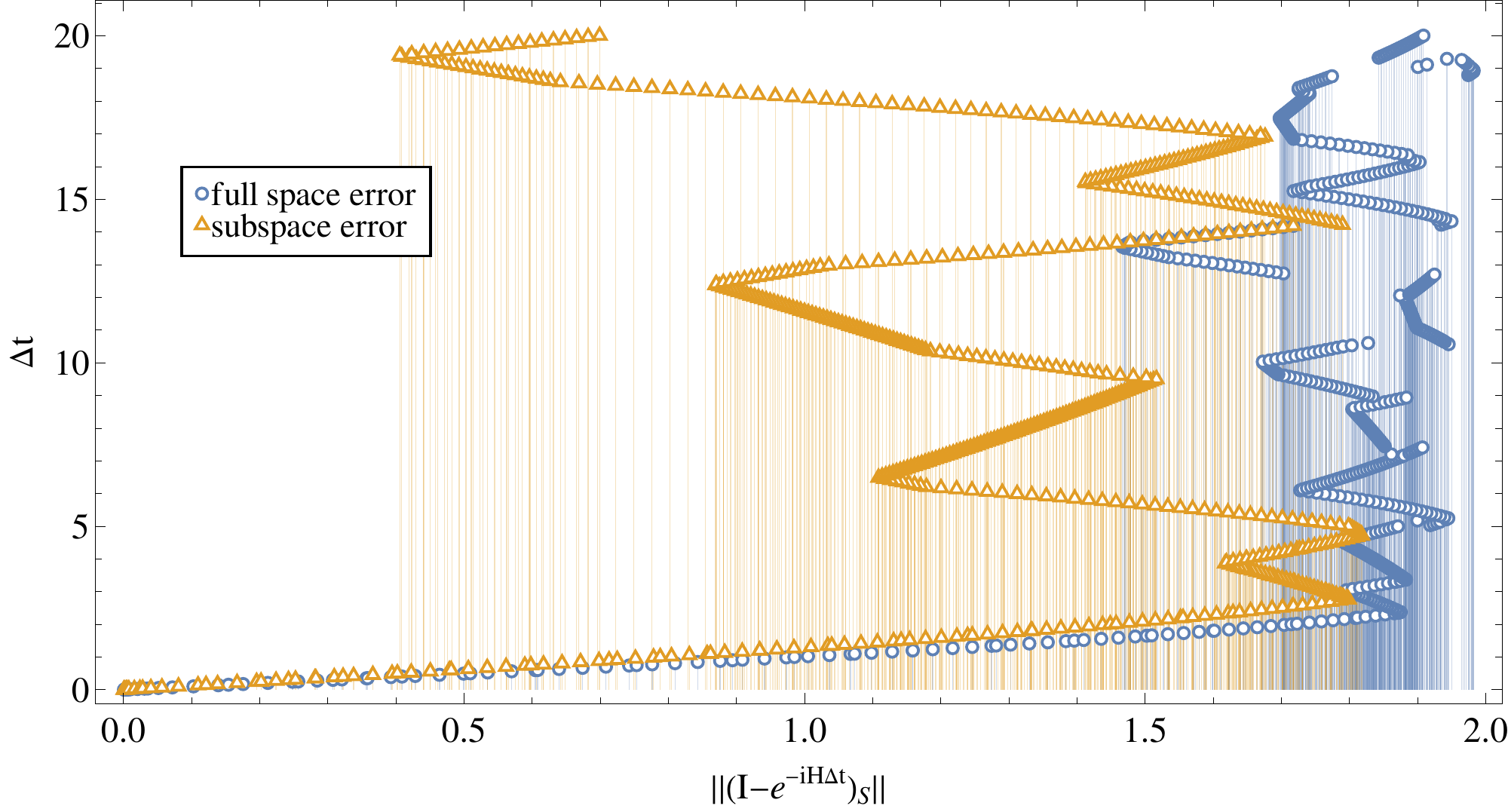}}
    \caption{\label{fig:trivial} The subspace and full space errors of propagators to the identity matrix
        with 598 different time intervals $\Delta t$ ranging from 0.001 to 20.
        The image of the full space errors are larger than the subspace error. This is 
        expected since the size of the subspace is 1/4 of the full space.
    }
\end{figure}

We synthesise circuits of the time-evolution operator of $\text H_2$ molecule
for various $\Delta t$ defined in \Cref{eq:deltat}. To justify that we are not
running some trivial problems, we compute ``non-triviality'', defined as the distance
--- for both subspace and full space ---  to the identity matrix as shown in
\Cref{fig:trivial}. The problem is less trivial for larger $\Delta t$.
From there, we decided to run various $\Delta t$ for up to 4.5 seconds.

\begin{figure}[hbt]
    \resizebox{.7\columnwidth}{!}{
    \includegraphics{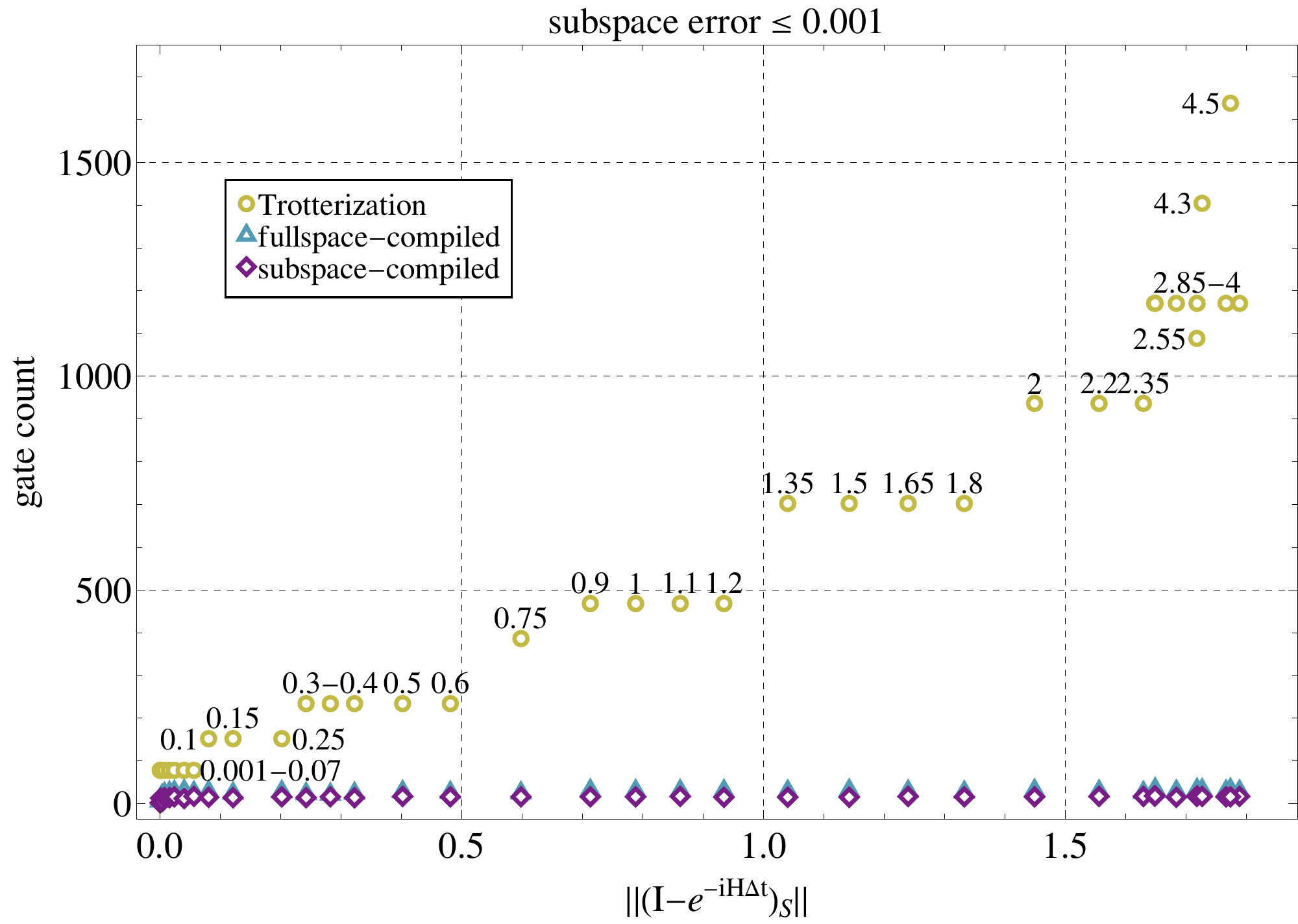}}
    \caption{\label{fig:subspace_trivial} The lowest gate count of successful results (subspace error $\leq 10^{-3}$)
        of the Trotterisations considered in 
        \Cref{fig:all_trotter}, subspace compilation, and full space compilation.
        The plot shows the count against non-triviality, \ie distance to the identity matrix in the subspace.
        The simulation duration $\Delta t$ shown on the corresponding marker of the Trotterisation counts to examine the 
        non-triviality (in the subspace) growth with respect to $\Delta t$. 
    }
\end{figure}

\begin{figure}[hbt]
    \resizebox{.7\columnwidth}{!}{
    \includegraphics{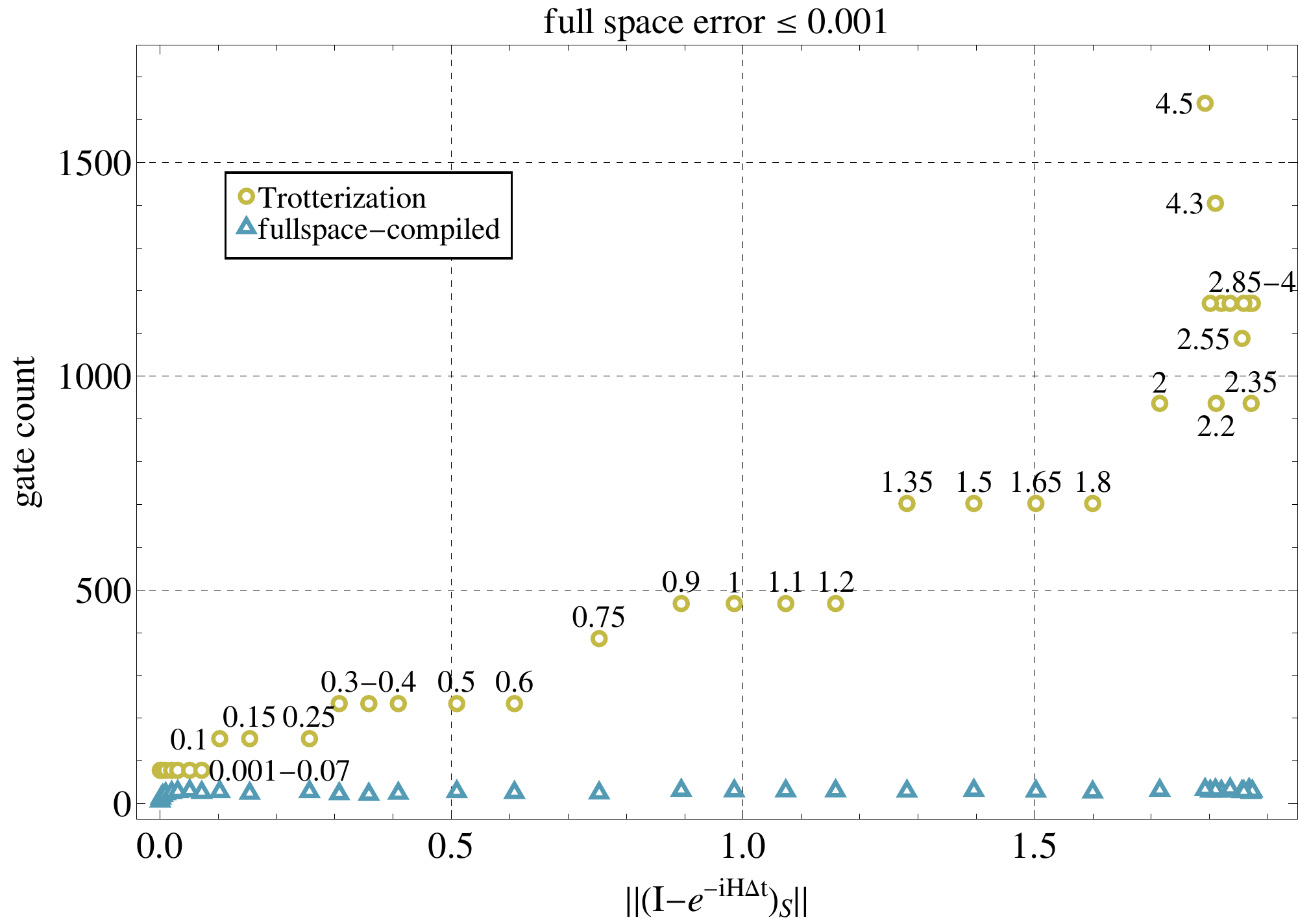}}
    \caption{\label{fig:fullspace_trivial} The lowest gate count of successful results (full space error $\leq 10^{-3}$)
        of the Trotterisations considered in 
        \Cref{fig:all_trotter} and full space compilation.
        The plot shows the count against non-triviality, \ie distance to the identity matrix in the full space.
        The simulation duration $\Delta t$ shown on the corresponding marker of the Trotterisation counts to examine the 
        non-triviality growth (in the full space) with respect to $\Delta t$. 
    }
\end{figure}

Plotting the best gate count for each cases as of \Cref{fig:dyn_main}, but
against the distance to the identity matrix, results in
\Cref{fig:subspace_trivial} and \Cref{fig:fullspace_trivial} for the subspace
and full space errors, respectively. These results show that the non-triviality
does not correlate with the gate count in the Trotterisation but rather with
the $\Delta t$. 

\begin{figure}[h]
    \includegraphics[width=\textwidth]{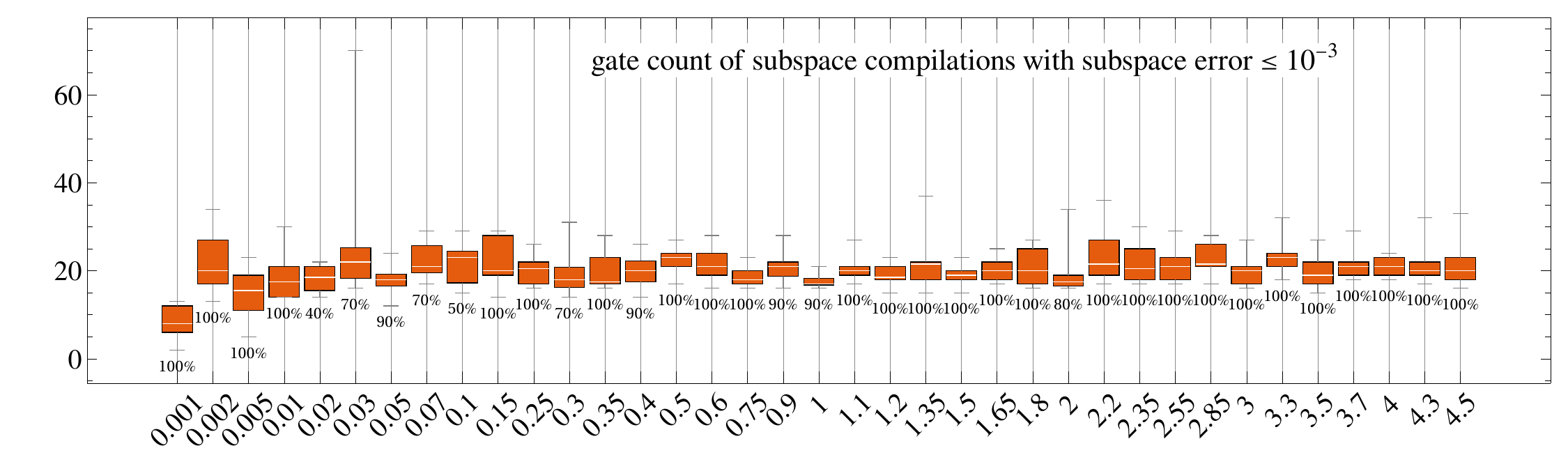}
    \includegraphics[width=\textwidth]{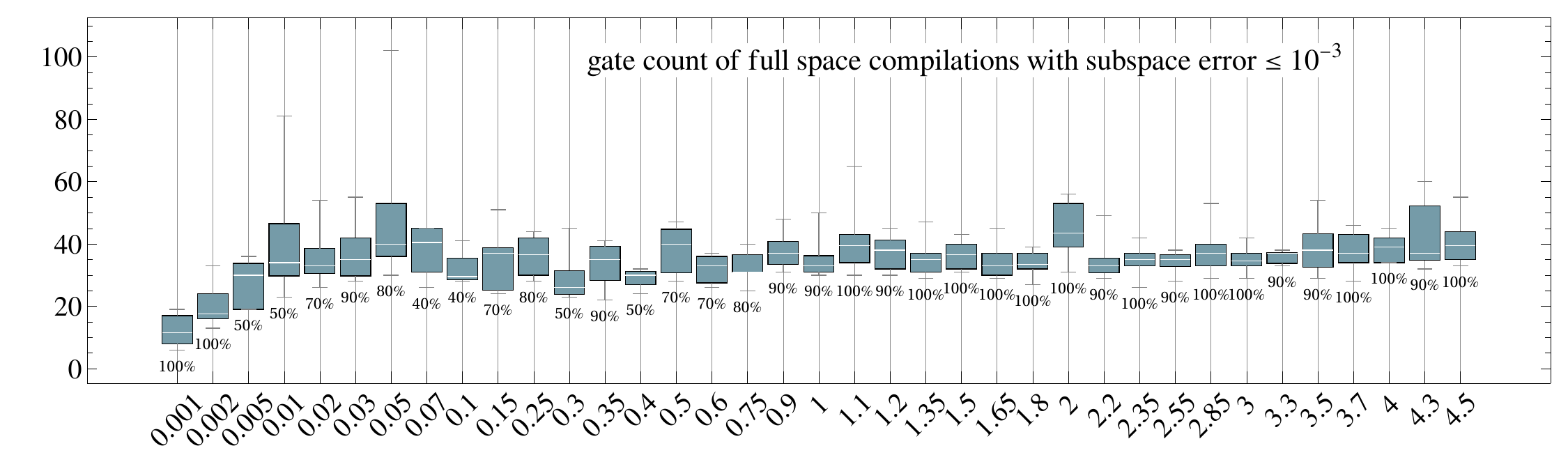}
    \includegraphics[width=\textwidth]{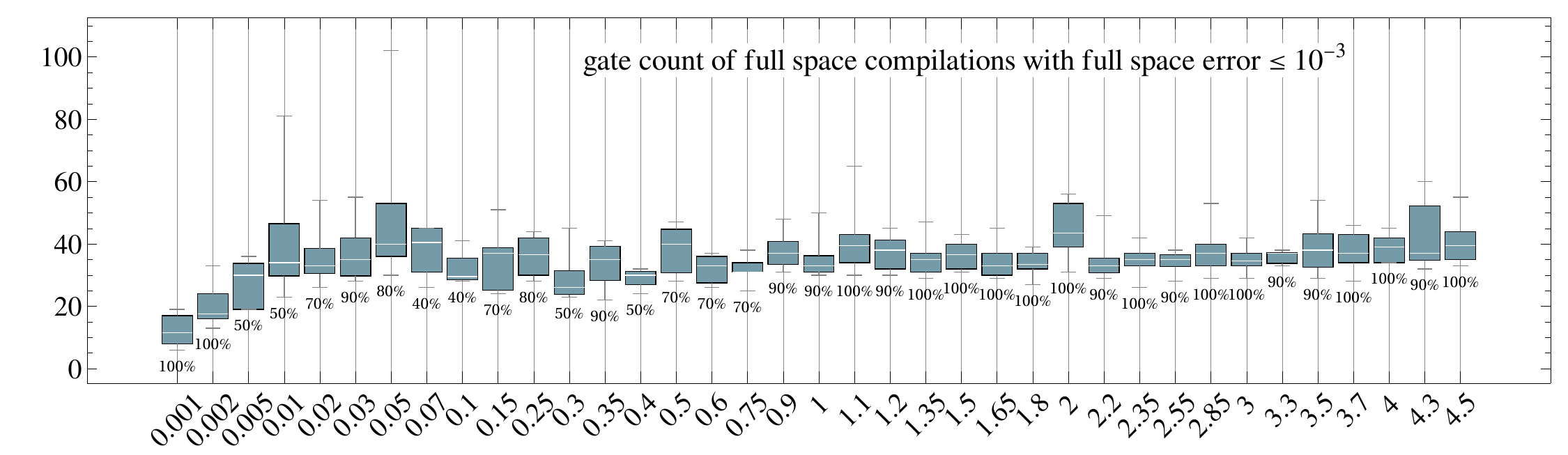}
    \caption{\label{fig:gatecountstat}
        Gate counts of successful results with two different successful criteria:
        subspace error $\leq 10^{-3}$ and full space error $\leq 10^{-3}$, for 
        the subspace and full space compilation. The numbers on $x$-axis denote time durations
        $\Delta t$ and the labels below whiskers signifying the success rate of ten trials.
    }
\end{figure}

\begin{figure}[ht]
    \resizebox{.9\columnwidth}{!}{
    \includegraphics{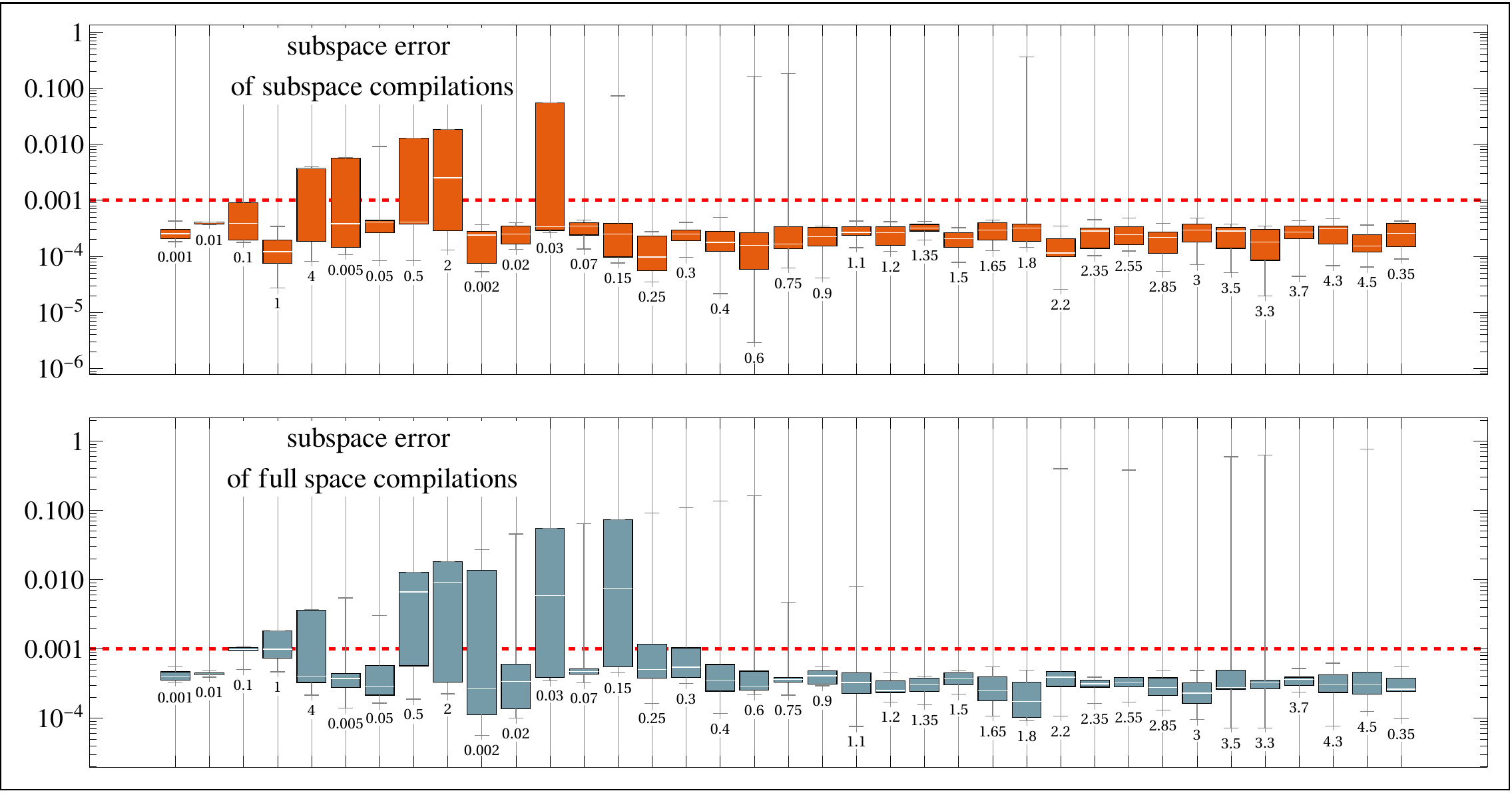}}
    \caption{\label{fig:subdists} Subspace errors. The red dashed lines indicate the goal error: $10^{-3}$.
        Both subspace and full space compilation show similar performance. 
    }
\end{figure}

\begin{figure}[ht]
    \resizebox{.9\columnwidth}{!}{
    \includegraphics{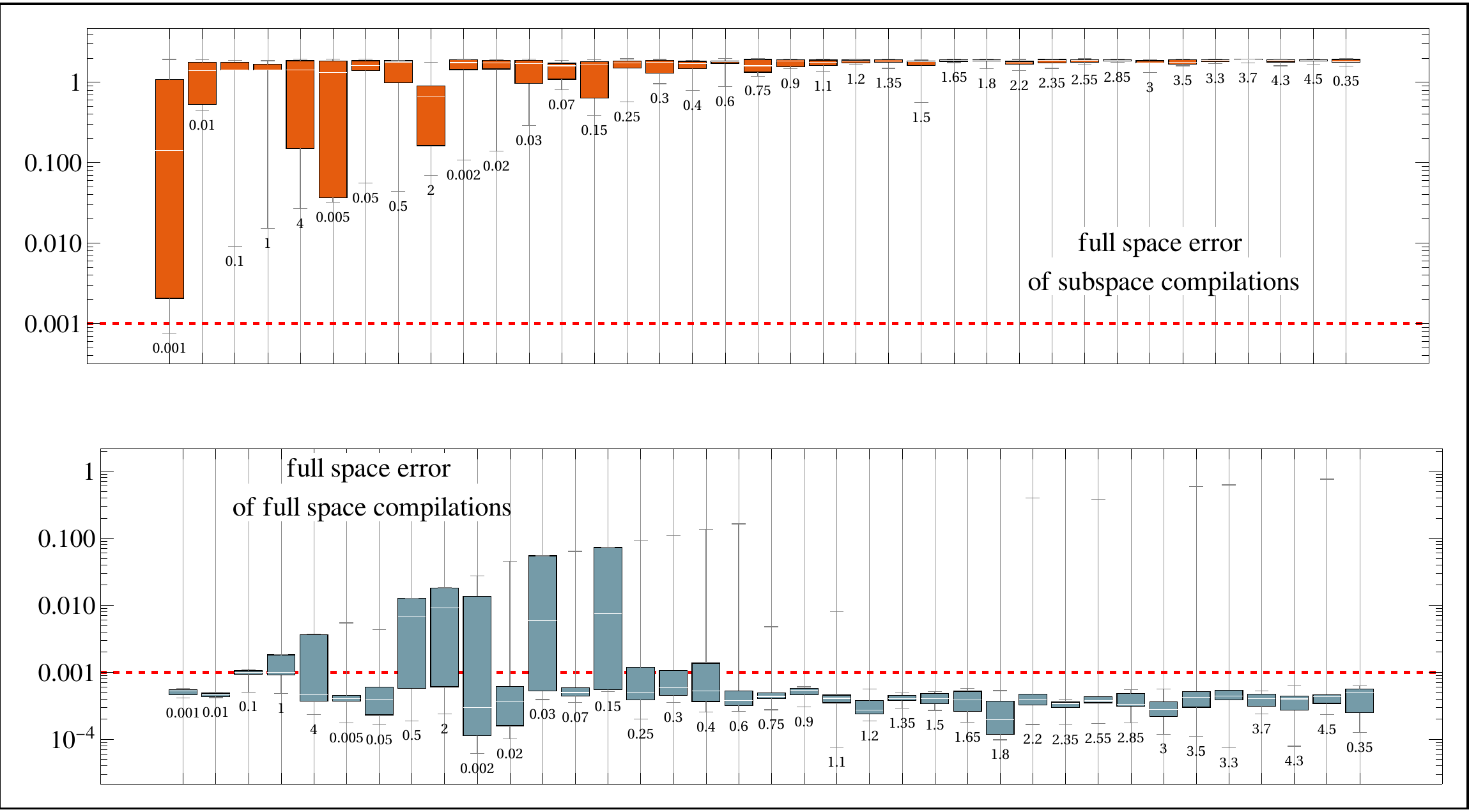}}
    \caption{\label{fig:fulldists} Full space errors. The red dashed lines indicate the goal error: $10^{-3}$.
        The subspace compilation is obviously perform poorly in this measure since it only takes into account the subspace. 
    }
\end{figure}

\begin{figure}[ht]
    \resizebox{.9\columnwidth}{!}{
    \includegraphics{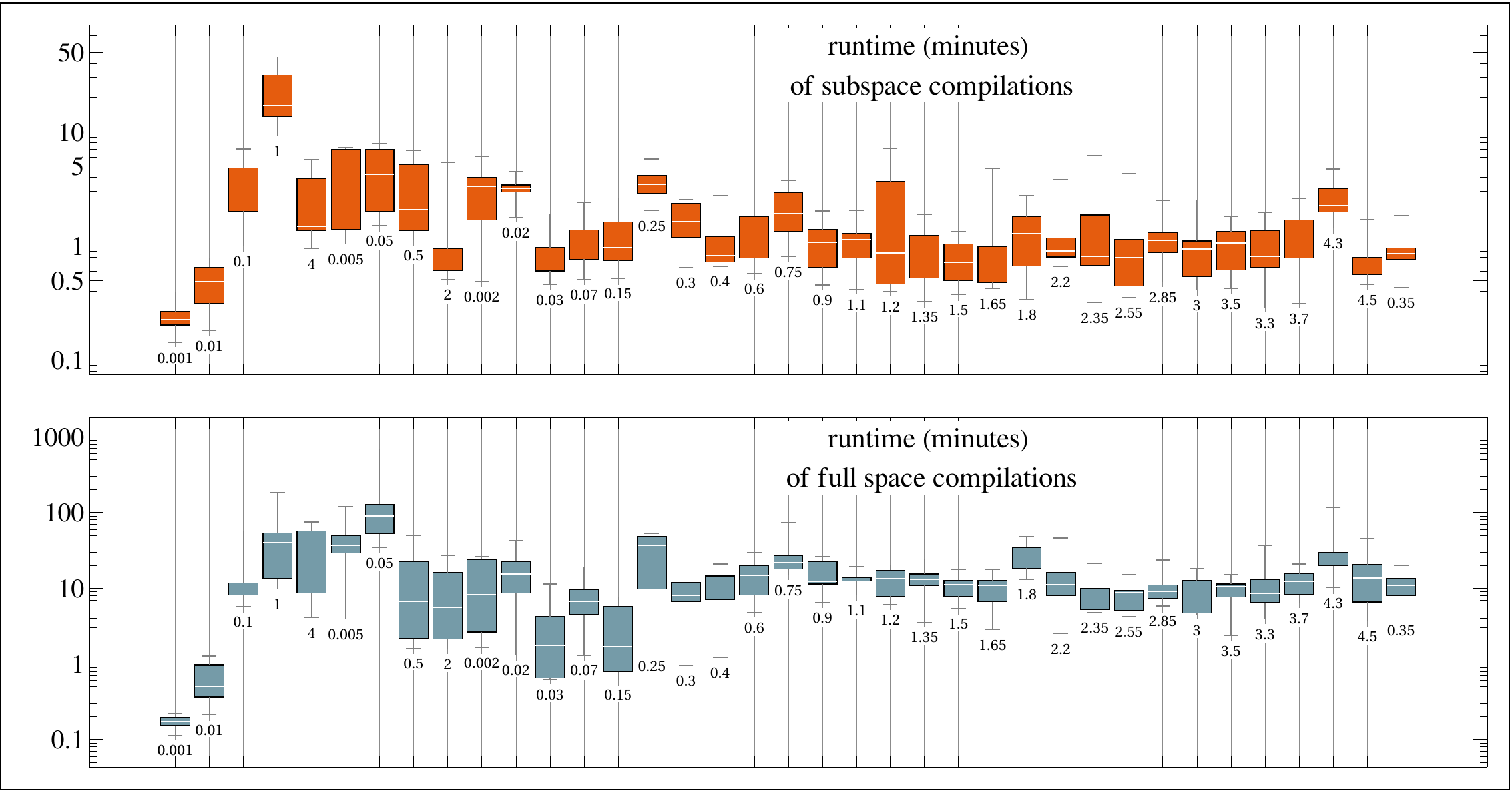}}
    \caption{\label{fig:runtimes} The total recompilation runtime for the subspace and full space compilation.
    }
\end{figure}

In synthesising circuits of the time-evolution operators, we run 10 trials of
circuit synthesis for each $\Delta t$, for the subspace and the full space.
The computational basis set of our Hydrogen problem is $\{\ket i\}$, for $i\in\{0,\dots
15\}$.
We use the least-significant bit notations, \eg $\ket{5}=\ket{0101}$. Thus, 
the subspace $\mathcal S$, the subspace that preserves the number of particles as two is spanned by 
\[\text{span}(\mathcal S)=\{\ket{0011},\ket{0101},\ket{0110},\ket{1001},\ket{1010},\ket{1100}\} \equiv \{\ket3, \ket5, \ket6, \ket9, \ket{10}, \ket{12}\} \].
With that, we obtain the statistic shown in \Cref{fig:gatecountstat,fig:subdists,fig:fulldists,fig:runtimes} from our compilations.

From \Cref{fig:gatecountstat}, the subspace compilations show lower gate count
with more successful rate compared to the full space compilations. The
successful results of full space compilations by criteria subspace and full
space distance are almost identical with only difference in result of $\Delta
t=0.75$.

From \Cref{fig:subdists}, we see both methods perform relatively similar.
However, it is interesting that for some small $\Delta t$, the optimiser shows
some struggle, in contrast to Trotterisation that perform consistently well for
small $\Delta t$ --- see \Cref{fig:all_trotter}.

The total optimisation runtime for all compilations shown in \Cref{fig:runtimes}; as we expected, subspace compilations
converge in shorter times than the full space compilations.

\end{document}

%% file: lih.tex
    \begin{tikzpicture}
        \node[]{\huge
    \begin{quantikz}[transparent,column sep={1.4cm,between origins},row sep={1.2cm,between origins}]
        \lstick{\Huge$\ket{1}_{0}\;$}&\qw&\qw&\qw&\qw&\qw&\qw&\qw&\qw&\qw&\qw&\gate[style={inner xsep=-3}]{\bm{\text{\bfseries Rz}_{3.1}}}&\qw&\qw&\qw&\qw&\qw&\qw\\
        \lstick{\Huge$\ket{1}_{1}\;$}&\qw&\qw&\qw&\qw&\qw&\qw\qw&\qw&\qw&\qw&\qw&\qw&\qw&\qw&\qw&\qw&\qw&\qw\\
        \lstick{\Huge$\ket{1}_{2}\;$}&\qw&\qw&\qw&\qw&\qw&\qw&\qw&\qw&\qw&\gate[style={inner xsep=-3}]{\bm{\text{\bfseries Rx}_{-\pi}}}&\ctrl{-2}\qw&\qw&\ctrl{1}\qw&\qw&\qw&\qw&\qw\\
        \lstick{\Huge$\ket{1}_{3}\;$}&\qw&\qw&\qw&\qw&\gate[style={inner xsep=-3}]{\bm{\text{\bfseries Rx}_{0.08}}}&\qw&\gate[style={inner xsep=-3}]{\bm{\text{\bfseries Rx}_{-0.07}}}&\qw&\ctrl{1}\qw&\qw&\gate[style={inner xsep=-3}]{\bm{\text{\bfseries Ry}_{\pi}}}&\qw&\gate[style={inner xsep=-3}]{\bm{\text{\bfseries Ry}_{-\pi}}}&\ctrl{2}\qw&\qw&\qw&\qw\\
        \lstick{\Huge$\ket{0}_{4}\;$}&\gate[style={inner xsep=-3}]{\bm{\text{\bfseries Ry}_{-0.43}}}&\qw&\ctrl{1}\qw&\qw&\qw&\qw&\qw&\qw&\gate[style={inner xsep=-3}]{\bm{\text{\bfseries Rx}_{-0.98}}}&\qw&\gate[style={inner xsep=-3}]{\bm{\text{\bfseries Rx}_{1.41}}}&\qw&\qw&\qw&\gate[style={inner xsep=-3}]{\bm{\text{\bfseries Rx}_{-1}}}&\ctrl{6}\qw&\qw\\
        \lstick{\Huge$\ket{0}_{5}\;$}&\gate[style={inner xsep=-3}]{\bm{\text{\bfseries Ry}_{0.26}}}&\ctrl{6}\qw&\gate[style={inner xsep=-3}]{\bm{\text{\bfseries Rz}_{\pi}}}&\ctrl{5}\qw&\ctrl{-2}\qw&\qw&\gate[style={inner xsep=-3}]{\bm{\text{\bfseries Rx}_\pi}}&\qw&\qw&\qw&\qw&\qw&\qw&\gate[style={inner xsep=-3}]{\bm{\text{\bfseries Rx}_{-\pi}}}&\ctrl{-1}\qw&\qw&\qw\\
        \lstick{\Huge$\ket{0}_{6}\;$}&\qw&\qw&\qw&\qw&\qw&\qw&\qw&\qw&\qw&\qw&\qw&\qw&\qw&\qw&\qw&\qw&\qw\\
        \lstick{\Huge$\ket{0}_{7}\;$}&\qw&\qw&\gate[style={inner xsep=-3}]{\bm{\text{\bfseries Rz}_{-2.21}}}&\qw&\qw&\qw&\qw&\qw&\qw&\qw&\qw&\qw&\qw&\qw&\qw&\qw&\qw\\
        \lstick{\Huge$\ket{0}_{8}\;$}&\qw&\qw&\qw&\qw&\qw&\qw&\qw&\qw&\qw&\qw&\qw&\qw&\qw&\qw&\qw&\qw&\qw\\
        \lstick{\Huge$\ket{0}_{9}\;$}&\qw&\qw&\qw&\qw&\qw&\qw&\qw&\qw&\qw&\qw&\qw&\qw&\qw&\qw&\qw&\qw&\qw\\
        \lstick{\Huge$\ket{0}_{10}$}&\qw&\qw&\qw&\gate[style={inner xsep=-3}]{\bm{\text{\bfseries Ry}_{3.01}}}&\qw&\gate[style={inner xsep=-3}]{\bm{\text{\bfseries Rz}_{2.01}}}&\qw&\gate[style={inner xsep=-3}]{\bm{\text{\bfseries Rx}_{-0.14}}}&\qw&\ctrl{-8}\qw&\qw&\qw&\qw&\qw&\qw&\gate[style={inner xsep=-3}]{\bm{\text{\bfseries Rx}_\pi}}&\qw\\
        \lstick{\Huge$\ket{0}_{11}$}&\qw&\gate[style={inner xsep=-3}]{\bm{\text{\bfseries Rx}_\pi}}&\ctrl{-4}\qw&\qw&\qw&\qw&\qw&\qw&\qw&\qw&\qw&\qw&\qw&\qw&\qw&\qw&\qw
\end{quantikz}
};
\end{tikzpicture}

%% file: main.bbl
\begin{thebibliography}{57}%
\makeatletter
\providecommand \@ifxundefined [1]{%
 \@ifx{#1\undefined}
}%
\providecommand \@ifnum [1]{%
 \ifnum #1\expandafter \@firstoftwo
 \else \expandafter \@secondoftwo
 \fi
}%
\providecommand \@ifx [1]{%
 \ifx #1\expandafter \@firstoftwo
 \else \expandafter \@secondoftwo
 \fi
}%
\providecommand \natexlab [1]{#1}%
\providecommand \enquote  [1]{``#1''}%
\providecommand \bibnamefont  [1]{#1}%
\providecommand \bibfnamefont [1]{#1}%
\providecommand \citenamefont [1]{#1}%
\providecommand \href@noop [0]{\@secondoftwo}%
\providecommand \href [0]{\begingroup \@sanitize@url \@href}%
\providecommand \@href[1]{\@@startlink{#1}\@@href}%
\providecommand \@@href[1]{\endgroup#1\@@endlink}%
\providecommand \@sanitize@url [0]{\catcode `\\12\catcode `\$12\catcode
  `\&12\catcode `\#12\catcode `\^12\catcode `\_12\catcode `\%12\relax}%
\providecommand \@@startlink[1]{}%
\providecommand \@@endlink[0]{}%
\providecommand \url  [0]{\begingroup\@sanitize@url \@url }%
\providecommand \@url [1]{\endgroup\@href {#1}{\urlprefix }}%
\providecommand \urlprefix  [0]{URL }%
\providecommand \Eprint [0]{\href }%
\providecommand \doibase [0]{http://dx.doi.org/}%
\providecommand \selectlanguage [0]{\@gobble}%
\providecommand \bibinfo  [0]{\@secondoftwo}%
\providecommand \bibfield  [0]{\@secondoftwo}%
\providecommand \translation [1]{[#1]}%
\providecommand \BibitemOpen [0]{}%
\providecommand \bibitemStop [0]{}%
\providecommand \bibitemNoStop [0]{.\EOS\space}%
\providecommand \EOS [0]{\spacefactor3000\relax}%
\providecommand \BibitemShut  [1]{\csname bibitem#1\endcsname}%
\let\auto@bib@innerbib\@empty
\bibitem [{\citenamefont {Preskill}(2018)}]{preskill2018quantum}%
  \BibitemOpen
  \bibfield  {author} {\bibinfo {author} {\bibfnamefont {John}\ \bibnamefont
  {Preskill}},\ }\bibfield  {title} {\enquote {\bibinfo {title} {Quantum
  computing in the nisq era and beyond},}\ }\href {\doibase
  10.22331/q-2018-08-06-79} {\bibfield  {journal} {\bibinfo  {journal}
  {Quantum}\ }\textbf {\bibinfo {volume} {2}},\ \bibinfo {pages} {79} (\bibinfo
  {year} {2018})}\BibitemShut {NoStop}%
\bibitem [{\citenamefont {Peruzzo}\ \emph {et~al.}(2014)\citenamefont
  {Peruzzo}, \citenamefont {McClean}, \citenamefont {Shadbolt}, \citenamefont
  {Yung}, \citenamefont {Zhou}, \citenamefont {Love}, \citenamefont
  {Aspuru-Guzik},\ and\ \citenamefont {O’brien}}]{peruzzo2014variational}%
  \BibitemOpen
  \bibfield  {author} {\bibinfo {author} {\bibfnamefont {Alberto}\ \bibnamefont
  {Peruzzo}}, \bibinfo {author} {\bibfnamefont {Jarrod}\ \bibnamefont
  {McClean}}, \bibinfo {author} {\bibfnamefont {Peter}\ \bibnamefont
  {Shadbolt}}, \bibinfo {author} {\bibfnamefont {Man-Hong}\ \bibnamefont
  {Yung}}, \bibinfo {author} {\bibfnamefont {Xiao-Qi}\ \bibnamefont {Zhou}},
  \bibinfo {author} {\bibfnamefont {Peter~J}\ \bibnamefont {Love}}, \bibinfo
  {author} {\bibfnamefont {Al{\'a}n}\ \bibnamefont {Aspuru-Guzik}}, \ and\
  \bibinfo {author} {\bibfnamefont {Jeremy~L}\ \bibnamefont {O’brien}},\
  }\bibfield  {title} {\enquote {\bibinfo {title} {A variational eigenvalue
  solver on a photonic quantum processor},}\ }\href {\doibase
  10.1038/ncomms5213} {\bibfield  {journal} {\bibinfo  {journal} {Nature
  communications}\ }\textbf {\bibinfo {volume} {5}},\ \bibinfo {pages} {1--7}
  (\bibinfo {year} {2014})}\BibitemShut {NoStop}%
\bibitem [{\citenamefont {Kempe}\ \emph {et~al.}(2005)\citenamefont {Kempe},
  \citenamefont {Kitaev},\ and\ \citenamefont {Regev}}]{kempe2004the}%
  \BibitemOpen
  \bibfield  {author} {\bibinfo {author} {\bibfnamefont {Julia}\ \bibnamefont
  {Kempe}}, \bibinfo {author} {\bibfnamefont {Alexei}\ \bibnamefont {Kitaev}},
  \ and\ \bibinfo {author} {\bibfnamefont {Oded}\ \bibnamefont {Regev}},\
  }\bibfield  {title} {\enquote {\bibinfo {title} {The complexity of the local
  hamiltonian problem},}\ }in\ \href@noop {} {\emph {\bibinfo {booktitle}
  {FSTTCS 2004: Foundations of Software Technology and Theoretical Computer
  Science}}},\ \bibinfo {editor} {edited by\ \bibinfo {editor} {\bibfnamefont
  {Kamal}\ \bibnamefont {Lodaya}}\ and\ \bibinfo {editor} {\bibfnamefont
  {Meena}\ \bibnamefont {Mahajan}}}\ (\bibinfo  {publisher} {Springer Berlin
  Heidelberg},\ \bibinfo {address} {Berlin, Heidelberg},\ \bibinfo {year}
  {2005})\ pp.\ \bibinfo {pages} {372--383}\BibitemShut {NoStop}%
\bibitem [{\citenamefont {Grimsley}\ \emph {et~al.}(2019)\citenamefont
  {Grimsley}, \citenamefont {Economou}, \citenamefont {Barnes},\ and\
  \citenamefont {Mayhall}}]{grimsley2019adaptive}%
  \BibitemOpen
  \bibfield  {author} {\bibinfo {author} {\bibfnamefont {Harper~R}\
  \bibnamefont {Grimsley}}, \bibinfo {author} {\bibfnamefont {Sophia~E}\
  \bibnamefont {Economou}}, \bibinfo {author} {\bibfnamefont {Edwin}\
  \bibnamefont {Barnes}}, \ and\ \bibinfo {author} {\bibfnamefont {Nicholas~J}\
  \bibnamefont {Mayhall}},\ }\bibfield  {title} {\enquote {\bibinfo {title} {An
  adaptive variational algorithm for exact molecular simulations on a quantum
  computer},}\ }\href {\doibase 10.1038/s41467-019-10988-2} {\bibfield
  {journal} {\bibinfo  {journal} {Nature communications}\ }\textbf {\bibinfo
  {volume} {10}},\ \bibinfo {pages} {1--9} (\bibinfo {year}
  {2019})}\BibitemShut {NoStop}%
\bibitem [{\citenamefont {Ollitrault}\ \emph {et~al.}(2020)\citenamefont
  {Ollitrault}, \citenamefont {Baiardi}, \citenamefont {Reiher},\ and\
  \citenamefont {Tavernelli}}]{ollitrault2020hardware}%
  \BibitemOpen
  \bibfield  {author} {\bibinfo {author} {\bibfnamefont {Pauline~J}\
  \bibnamefont {Ollitrault}}, \bibinfo {author} {\bibfnamefont {Alberto}\
  \bibnamefont {Baiardi}}, \bibinfo {author} {\bibfnamefont {Markus}\
  \bibnamefont {Reiher}}, \ and\ \bibinfo {author} {\bibfnamefont {Ivano}\
  \bibnamefont {Tavernelli}},\ }\bibfield  {title} {\enquote {\bibinfo {title}
  {Hardware efficient quantum algorithms for vibrational structure
  calculations},}\ }\href {\doibase 10.1039/D0SC01908A} {\bibfield  {journal}
  {\bibinfo  {journal} {Chemical science}\ }\textbf {\bibinfo {volume} {11}},\
  \bibinfo {pages} {6842--6855} (\bibinfo {year} {2020})}\BibitemShut {NoStop}%
\bibitem [{\citenamefont {Delgado}\ \emph {et~al.}(2021)\citenamefont
  {Delgado}, \citenamefont {Arrazola}, \citenamefont {Jahangiri}, \citenamefont
  {Niu}, \citenamefont {Izaac}, \citenamefont {Roberts},\ and\ \citenamefont
  {Killoran}}]{delgado2021variational}%
  \BibitemOpen
  \bibfield  {author} {\bibinfo {author} {\bibfnamefont {Alain}\ \bibnamefont
  {Delgado}}, \bibinfo {author} {\bibfnamefont {Juan~Miguel}\ \bibnamefont
  {Arrazola}}, \bibinfo {author} {\bibfnamefont {Soran}\ \bibnamefont
  {Jahangiri}}, \bibinfo {author} {\bibfnamefont {Zeyue}\ \bibnamefont {Niu}},
  \bibinfo {author} {\bibfnamefont {Josh}\ \bibnamefont {Izaac}}, \bibinfo
  {author} {\bibfnamefont {Chase}\ \bibnamefont {Roberts}}, \ and\ \bibinfo
  {author} {\bibfnamefont {Nathan}\ \bibnamefont {Killoran}},\ }\bibfield
  {title} {\enquote {\bibinfo {title} {Variational quantum algorithm for
  molecular geometry optimization},}\ }\href {\doibase
  10.1103/PhysRevA.104.052402} {\bibfield  {journal} {\bibinfo  {journal}
  {Physical Review A}\ }\textbf {\bibinfo {volume} {104}},\ \bibinfo {pages}
  {052402} (\bibinfo {year} {2021})}\BibitemShut {NoStop}%
\bibitem [{\citenamefont {Gibbs}\ \emph {et~al.}(2021)\citenamefont {Gibbs},
  \citenamefont {Gili}, \citenamefont {Holmes}, \citenamefont {Commeau},
  \citenamefont {Arrasmith}, \citenamefont {Cincio}, \citenamefont {Coles},\
  and\ \citenamefont {Sornborger}}]{gibbs2021long}%
  \BibitemOpen
  \bibfield  {author} {\bibinfo {author} {\bibfnamefont {Joe}\ \bibnamefont
  {Gibbs}}, \bibinfo {author} {\bibfnamefont {Kaitlin}\ \bibnamefont {Gili}},
  \bibinfo {author} {\bibfnamefont {Zo{\"e}}\ \bibnamefont {Holmes}}, \bibinfo
  {author} {\bibfnamefont {Benjamin}\ \bibnamefont {Commeau}}, \bibinfo
  {author} {\bibfnamefont {Andrew}\ \bibnamefont {Arrasmith}}, \bibinfo
  {author} {\bibfnamefont {Lukasz}\ \bibnamefont {Cincio}}, \bibinfo {author}
  {\bibfnamefont {Patrick~J}\ \bibnamefont {Coles}}, \ and\ \bibinfo {author}
  {\bibfnamefont {Andrew}\ \bibnamefont {Sornborger}},\ }\bibfield  {title}
  {\enquote {\bibinfo {title} {Long-time simulations with high fidelity on
  quantum hardware},}\ }\href {\doibase 10.48550/arXiv.2102.04313} {\bibfield
  {journal} {\bibinfo  {journal} {arXiv preprint arXiv:2102.04313}\ } (\bibinfo
  {year} {2021}),\ 10.48550/arXiv.2102.04313}\BibitemShut {NoStop}%
\bibitem [{\citenamefont {Greene-Diniz}\ and\ \citenamefont
  {Mu{\~n}oz~Ramo}(2021)}]{greene2021generalized}%
  \BibitemOpen
  \bibfield  {author} {\bibinfo {author} {\bibfnamefont {Gabriel}\ \bibnamefont
  {Greene-Diniz}}\ and\ \bibinfo {author} {\bibfnamefont {David}\ \bibnamefont
  {Mu{\~n}oz~Ramo}},\ }\bibfield  {title} {\enquote {\bibinfo {title}
  {Generalized unitary coupled cluster excitations for multireference molecular
  states optimized by the variational quantum eigensolver},}\ }\href {\doibase
  10.1002/qua.26352} {\bibfield  {journal} {\bibinfo  {journal} {International
  Journal of Quantum Chemistry}\ }\textbf {\bibinfo {volume} {121}},\ \bibinfo
  {pages} {e26352} (\bibinfo {year} {2021})}\BibitemShut {NoStop}%
\bibitem [{\citenamefont {Metcalf}\ \emph {et~al.}(2020)\citenamefont
  {Metcalf}, \citenamefont {Bauman}, \citenamefont {Kowalski},\ and\
  \citenamefont {De~Jong}}]{metcalf2020resource}%
  \BibitemOpen
  \bibfield  {author} {\bibinfo {author} {\bibfnamefont {Mekena}\ \bibnamefont
  {Metcalf}}, \bibinfo {author} {\bibfnamefont {Nicholas~P}\ \bibnamefont
  {Bauman}}, \bibinfo {author} {\bibfnamefont {Karol}\ \bibnamefont
  {Kowalski}}, \ and\ \bibinfo {author} {\bibfnamefont {Wibe~A}\ \bibnamefont
  {De~Jong}},\ }\bibfield  {title} {\enquote {\bibinfo {title}
  {Resource-efficient chemistry on quantum computers with the variational
  quantum eigensolver and the double unitary coupled-cluster approach},}\
  }\href {\doibase 10.1021/acs.jctc.0c00421} {\bibfield  {journal} {\bibinfo
  {journal} {Journal of chemical theory and computation}\ }\textbf {\bibinfo
  {volume} {16}},\ \bibinfo {pages} {6165--6175} (\bibinfo {year}
  {2020})}\BibitemShut {NoStop}%
\bibitem [{\citenamefont {Chan}\ \emph {et~al.}(2021)\citenamefont {Chan},
  \citenamefont {Fitzpatrick}, \citenamefont {Segarra-Mart{\'\i}},
  \citenamefont {Bearpark},\ and\ \citenamefont {Tew}}]{chan2021molecular}%
  \BibitemOpen
  \bibfield  {author} {\bibinfo {author} {\bibfnamefont {Hans Hon~Sang}\
  \bibnamefont {Chan}}, \bibinfo {author} {\bibfnamefont {Nathan}\ \bibnamefont
  {Fitzpatrick}}, \bibinfo {author} {\bibfnamefont {Javier}\ \bibnamefont
  {Segarra-Mart{\'\i}}}, \bibinfo {author} {\bibfnamefont {Michael~J}\
  \bibnamefont {Bearpark}}, \ and\ \bibinfo {author} {\bibfnamefont {David~P}\
  \bibnamefont {Tew}},\ }\bibfield  {title} {\enquote {\bibinfo {title}
  {Molecular excited state calculations with adaptive wavefunctions on a
  quantum eigensolver emulation: reducing circuit depth and separating spin
  states},}\ }\href {\doibase 10.1039/D1CP02227J} {\bibfield  {journal}
  {\bibinfo  {journal} {Physical Chemistry Chemical Physics}\ }\textbf
  {\bibinfo {volume} {23}},\ \bibinfo {pages} {26438--26450} (\bibinfo {year}
  {2021})}\BibitemShut {NoStop}%
\bibitem [{\citenamefont {McArdle}\ \emph {et~al.}(2020)\citenamefont
  {McArdle}, \citenamefont {Endo}, \citenamefont {Aspuru-Guzik}, \citenamefont
  {Benjamin},\ and\ \citenamefont {Yuan}}]{mcardle2020quantum}%
  \BibitemOpen
  \bibfield  {author} {\bibinfo {author} {\bibfnamefont {Sam}\ \bibnamefont
  {McArdle}}, \bibinfo {author} {\bibfnamefont {Suguru}\ \bibnamefont {Endo}},
  \bibinfo {author} {\bibfnamefont {Al{\'a}n}\ \bibnamefont {Aspuru-Guzik}},
  \bibinfo {author} {\bibfnamefont {Simon~C}\ \bibnamefont {Benjamin}}, \ and\
  \bibinfo {author} {\bibfnamefont {Xiao}\ \bibnamefont {Yuan}},\ }\bibfield
  {title} {\enquote {\bibinfo {title} {Quantum computational chemistry},}\
  }\href {\doibase 10.1103/RevModPhys.92.015003} {\bibfield  {journal}
  {\bibinfo  {journal} {Reviews of Modern Physics}\ }\textbf {\bibinfo {volume}
  {92}},\ \bibinfo {pages} {015003} (\bibinfo {year} {2020})}\BibitemShut
  {NoStop}%
\bibitem [{\citenamefont {Li}\ \emph {et~al.}(2019)\citenamefont {Li},
  \citenamefont {Hu}, \citenamefont {Zhang}, \citenamefont {Song},\ and\
  \citenamefont {Yung}}]{li2019variational}%
  \BibitemOpen
  \bibfield  {author} {\bibinfo {author} {\bibfnamefont {Yifan}\ \bibnamefont
  {Li}}, \bibinfo {author} {\bibfnamefont {Jiaqi}\ \bibnamefont {Hu}}, \bibinfo
  {author} {\bibfnamefont {Xiao-Ming}\ \bibnamefont {Zhang}}, \bibinfo {author}
  {\bibfnamefont {Zhigang}\ \bibnamefont {Song}}, \ and\ \bibinfo {author}
  {\bibfnamefont {Man-Hong}\ \bibnamefont {Yung}},\ }\bibfield  {title}
  {\enquote {\bibinfo {title} {Variational quantum simulation for quantum
  chemistry},}\ }\href {\doibase 10.1002/adts.201800182} {\bibfield  {journal}
  {\bibinfo  {journal} {Advanced Theory and Simulations}\ }\textbf {\bibinfo
  {volume} {2}},\ \bibinfo {pages} {1800182} (\bibinfo {year}
  {2019})}\BibitemShut {NoStop}%
\bibitem [{\citenamefont {Xu}\ \emph {et~al.}(2021)\citenamefont {Xu},
  \citenamefont {Sun}, \citenamefont {Endo}, \citenamefont {Li}, \citenamefont
  {Benjamin},\ and\ \citenamefont {Yuan}}]{xu2021variational}%
  \BibitemOpen
  \bibfield  {author} {\bibinfo {author} {\bibfnamefont {Xiaosi}\ \bibnamefont
  {Xu}}, \bibinfo {author} {\bibfnamefont {Jinzhao}\ \bibnamefont {Sun}},
  \bibinfo {author} {\bibfnamefont {Suguru}\ \bibnamefont {Endo}}, \bibinfo
  {author} {\bibfnamefont {Ying}\ \bibnamefont {Li}}, \bibinfo {author}
  {\bibfnamefont {Simon~C}\ \bibnamefont {Benjamin}}, \ and\ \bibinfo {author}
  {\bibfnamefont {Xiao}\ \bibnamefont {Yuan}},\ }\bibfield  {title} {\enquote
  {\bibinfo {title} {Variational algorithms for linear algebra},}\ }\href
  {\doibase 10.1016/j.scib.2021.06.023} {\bibfield  {journal} {\bibinfo
  {journal} {Science Bulletin}\ }\textbf {\bibinfo {volume} {66}},\ \bibinfo
  {pages} {2181--2188} (\bibinfo {year} {2021})}\BibitemShut {NoStop}%
\bibitem [{\citenamefont {Endo}\ \emph {et~al.}(2020)\citenamefont {Endo},
  \citenamefont {Sun}, \citenamefont {Li}, \citenamefont {Benjamin},\ and\
  \citenamefont {Yuan}}]{endo2020variational}%
  \BibitemOpen
  \bibfield  {author} {\bibinfo {author} {\bibfnamefont {Suguru}\ \bibnamefont
  {Endo}}, \bibinfo {author} {\bibfnamefont {Jinzhao}\ \bibnamefont {Sun}},
  \bibinfo {author} {\bibfnamefont {Ying}\ \bibnamefont {Li}}, \bibinfo
  {author} {\bibfnamefont {Simon~C}\ \bibnamefont {Benjamin}}, \ and\ \bibinfo
  {author} {\bibfnamefont {Xiao}\ \bibnamefont {Yuan}},\ }\bibfield  {title}
  {\enquote {\bibinfo {title} {Variational quantum simulation of general
  processes},}\ }\href {\doibase 10.1103/PhysRevLett.125.010501} {\bibfield
  {journal} {\bibinfo  {journal} {Physical Review Letters}\ }\textbf {\bibinfo
  {volume} {125}},\ \bibinfo {pages} {010501} (\bibinfo {year}
  {2020})}\BibitemShut {NoStop}%
\bibitem [{\citenamefont {Patil}\ \emph {et~al.}(2022)\citenamefont {Patil},
  \citenamefont {Wang},\ and\ \citenamefont
  {Krsti{\'c}}}]{patil2022variational}%
  \BibitemOpen
  \bibfield  {author} {\bibinfo {author} {\bibfnamefont {Hrushikesh}\
  \bibnamefont {Patil}}, \bibinfo {author} {\bibfnamefont {Yulun}\ \bibnamefont
  {Wang}}, \ and\ \bibinfo {author} {\bibfnamefont {Predrag~S}\ \bibnamefont
  {Krsti{\'c}}},\ }\bibfield  {title} {\enquote {\bibinfo {title} {Variational
  quantum linear solver with a dynamic ansatz},}\ }\href {\doibase
  10.1103/PhysRevA.105.012423} {\bibfield  {journal} {\bibinfo  {journal}
  {Physical Review A}\ }\textbf {\bibinfo {volume} {105}},\ \bibinfo {pages}
  {012423} (\bibinfo {year} {2022})}\BibitemShut {NoStop}%
\bibitem [{\citenamefont {Liu}\ \emph {et~al.}(2021)\citenamefont {Liu},
  \citenamefont {Wu}, \citenamefont {Wan}, \citenamefont {Pan}, \citenamefont
  {Qin}, \citenamefont {Gao},\ and\ \citenamefont {Wen}}]{liu2021variational}%
  \BibitemOpen
  \bibfield  {author} {\bibinfo {author} {\bibfnamefont {Hai-Ling}\
  \bibnamefont {Liu}}, \bibinfo {author} {\bibfnamefont {Yu-Sen}\ \bibnamefont
  {Wu}}, \bibinfo {author} {\bibfnamefont {Lin-Chun}\ \bibnamefont {Wan}},
  \bibinfo {author} {\bibfnamefont {Shi-Jie}\ \bibnamefont {Pan}}, \bibinfo
  {author} {\bibfnamefont {Su-Juan}\ \bibnamefont {Qin}}, \bibinfo {author}
  {\bibfnamefont {Fei}\ \bibnamefont {Gao}}, \ and\ \bibinfo {author}
  {\bibfnamefont {Qiao-Yan}\ \bibnamefont {Wen}},\ }\bibfield  {title}
  {\enquote {\bibinfo {title} {Variational quantum algorithm for the poisson
  equation},}\ }\href {\doibase 10.1103/PhysRevA.104.022418} {\bibfield
  {journal} {\bibinfo  {journal} {Physical Review A}\ }\textbf {\bibinfo
  {volume} {104}},\ \bibinfo {pages} {022418} (\bibinfo {year}
  {2021})}\BibitemShut {NoStop}%
\bibitem [{\citenamefont {Khatri}\ \emph {et~al.}(2019)\citenamefont {Khatri},
  \citenamefont {LaRose}, \citenamefont {Poremba}, \citenamefont {Cincio},
  \citenamefont {Sornborger},\ and\ \citenamefont {Coles}}]{khatri2019quantum}%
  \BibitemOpen
  \bibfield  {author} {\bibinfo {author} {\bibfnamefont {Sumeet}\ \bibnamefont
  {Khatri}}, \bibinfo {author} {\bibfnamefont {Ryan}\ \bibnamefont {LaRose}},
  \bibinfo {author} {\bibfnamefont {Alexander}\ \bibnamefont {Poremba}},
  \bibinfo {author} {\bibfnamefont {Lukasz}\ \bibnamefont {Cincio}}, \bibinfo
  {author} {\bibfnamefont {Andrew~T}\ \bibnamefont {Sornborger}}, \ and\
  \bibinfo {author} {\bibfnamefont {Patrick~J}\ \bibnamefont {Coles}},\
  }\bibfield  {title} {\enquote {\bibinfo {title} {Quantum-assisted quantum
  compiling},}\ }\href {\doibase doi.org/10.22331/q-2019-05-13-140} {\bibfield
  {journal} {\bibinfo  {journal} {Quantum}\ }\textbf {\bibinfo {volume} {3}},\
  \bibinfo {pages} {140} (\bibinfo {year} {2019})}\BibitemShut {NoStop}%
\bibitem [{\citenamefont {Jones}\ and\ \citenamefont
  {Benjamin}(2022)}]{jones2022robust}%
  \BibitemOpen
  \bibfield  {author} {\bibinfo {author} {\bibfnamefont {Tyson}\ \bibnamefont
  {Jones}}\ and\ \bibinfo {author} {\bibfnamefont {Simon~C}\ \bibnamefont
  {Benjamin}},\ }\bibfield  {title} {\enquote {\bibinfo {title} {Robust quantum
  compilation and circuit optimisation via energy minimisation},}\ }\href
  {\doibase 10.22331/q-2022-01-24-628} {\bibfield  {journal} {\bibinfo
  {journal} {Quantum}\ }\textbf {\bibinfo {volume} {6}},\ \bibinfo {pages}
  {628} (\bibinfo {year} {2022})}\BibitemShut {NoStop}%
\bibitem [{\citenamefont {Gokhale}\ \emph {et~al.}(2019)\citenamefont
  {Gokhale}, \citenamefont {Ding}, \citenamefont {Propson}, \citenamefont
  {Winkler}, \citenamefont {Leung}, \citenamefont {Shi}, \citenamefont
  {Schuster}, \citenamefont {Hoffmann},\ and\ \citenamefont
  {Chong}}]{gokhale2019partial}%
  \BibitemOpen
  \bibfield  {author} {\bibinfo {author} {\bibfnamefont {Pranav}\ \bibnamefont
  {Gokhale}}, \bibinfo {author} {\bibfnamefont {Yongshan}\ \bibnamefont
  {Ding}}, \bibinfo {author} {\bibfnamefont {Thomas}\ \bibnamefont {Propson}},
  \bibinfo {author} {\bibfnamefont {Christopher}\ \bibnamefont {Winkler}},
  \bibinfo {author} {\bibfnamefont {Nelson}\ \bibnamefont {Leung}}, \bibinfo
  {author} {\bibfnamefont {Yunong}\ \bibnamefont {Shi}}, \bibinfo {author}
  {\bibfnamefont {David~I}\ \bibnamefont {Schuster}}, \bibinfo {author}
  {\bibfnamefont {Henry}\ \bibnamefont {Hoffmann}}, \ and\ \bibinfo {author}
  {\bibfnamefont {Frederic~T}\ \bibnamefont {Chong}},\ }\bibfield  {title}
  {\enquote {\bibinfo {title} {Partial compilation of variational algorithms
  for noisy intermediate-scale quantum machines},}\ }in\ \href {\doibase
  10.1145/3352460.3358313} {\emph {\bibinfo {booktitle} {Proceedings of the
  52nd Annual IEEE/ACM International Symposium on Microarchitecture}}}\
  (\bibinfo {year} {2019})\ pp.\ \bibinfo {pages} {266--278}\BibitemShut
  {NoStop}%
\bibitem [{\citenamefont {Caro}\ \emph {et~al.}(2021)\citenamefont {Caro},
  \citenamefont {Huang}, \citenamefont {Cerezo}, \citenamefont {Sharma},
  \citenamefont {Sornborger}, \citenamefont {Cincio},\ and\ \citenamefont
  {Coles}}]{caro2021generalization}%
  \BibitemOpen
  \bibfield  {author} {\bibinfo {author} {\bibfnamefont {Matthias~C}\
  \bibnamefont {Caro}}, \bibinfo {author} {\bibfnamefont {Hsin-Yuan}\
  \bibnamefont {Huang}}, \bibinfo {author} {\bibfnamefont {M}~\bibnamefont
  {Cerezo}}, \bibinfo {author} {\bibfnamefont {Kunal}\ \bibnamefont {Sharma}},
  \bibinfo {author} {\bibfnamefont {Andrew}\ \bibnamefont {Sornborger}},
  \bibinfo {author} {\bibfnamefont {Lukasz}\ \bibnamefont {Cincio}}, \ and\
  \bibinfo {author} {\bibfnamefont {Patrick~J}\ \bibnamefont {Coles}},\
  }\bibfield  {title} {\enquote {\bibinfo {title} {Generalization in quantum
  machine learning from few training data},}\ }\href {\doibase
  10.48550/arXiv.2111.05292} {\bibfield  {journal} {\bibinfo  {journal} {arXiv
  preprint arXiv:2111.05292}\ } (\bibinfo {year} {2021}),\
  10.48550/arXiv.2111.05292}\BibitemShut {NoStop}%
\bibitem [{\citenamefont {Cerezo}\ \emph {et~al.}(2021)\citenamefont {Cerezo},
  \citenamefont {Arrasmith}, \citenamefont {Babbush}, \citenamefont {Benjamin},
  \citenamefont {Endo}, \citenamefont {Fujii}, \citenamefont {McClean},
  \citenamefont {Mitarai}, \citenamefont {Yuan}, \citenamefont {Cincio} \emph
  {et~al.}}]{cerezo2021variational}%
  \BibitemOpen
  \bibfield  {author} {\bibinfo {author} {\bibfnamefont {Marco}\ \bibnamefont
  {Cerezo}}, \bibinfo {author} {\bibfnamefont {Andrew}\ \bibnamefont
  {Arrasmith}}, \bibinfo {author} {\bibfnamefont {Ryan}\ \bibnamefont
  {Babbush}}, \bibinfo {author} {\bibfnamefont {Simon~C}\ \bibnamefont
  {Benjamin}}, \bibinfo {author} {\bibfnamefont {Suguru}\ \bibnamefont {Endo}},
  \bibinfo {author} {\bibfnamefont {Keisuke}\ \bibnamefont {Fujii}}, \bibinfo
  {author} {\bibfnamefont {Jarrod~R}\ \bibnamefont {McClean}}, \bibinfo
  {author} {\bibfnamefont {Kosuke}\ \bibnamefont {Mitarai}}, \bibinfo {author}
  {\bibfnamefont {Xiao}\ \bibnamefont {Yuan}}, \bibinfo {author} {\bibfnamefont
  {Lukasz}\ \bibnamefont {Cincio}},  \emph {et~al.},\ }\bibfield  {title}
  {\enquote {\bibinfo {title} {Variational quantum algorithms},}\ }\href
  {\doibase 10.1038/s42254-021-00348-9} {\bibfield  {journal} {\bibinfo
  {journal} {Nature Reviews Physics}\ }\textbf {\bibinfo {volume} {3}},\
  \bibinfo {pages} {625--644} (\bibinfo {year} {2021})}\BibitemShut {NoStop}%
\bibitem [{\citenamefont {Tilly}\ \emph {et~al.}(2021)\citenamefont {Tilly},
  \citenamefont {Chen}, \citenamefont {Cao}, \citenamefont {Picozzi},
  \citenamefont {Setia}, \citenamefont {Li}, \citenamefont {Grant},
  \citenamefont {Wossnig}, \citenamefont {Rungger}, \citenamefont {Booth} \emph
  {et~al.}}]{tilly2021variational}%
  \BibitemOpen
  \bibfield  {author} {\bibinfo {author} {\bibfnamefont {Jules}\ \bibnamefont
  {Tilly}}, \bibinfo {author} {\bibfnamefont {Hongxiang}\ \bibnamefont {Chen}},
  \bibinfo {author} {\bibfnamefont {Shuxiang}\ \bibnamefont {Cao}}, \bibinfo
  {author} {\bibfnamefont {Dario}\ \bibnamefont {Picozzi}}, \bibinfo {author}
  {\bibfnamefont {Kanav}\ \bibnamefont {Setia}}, \bibinfo {author}
  {\bibfnamefont {Ying}\ \bibnamefont {Li}}, \bibinfo {author} {\bibfnamefont
  {Edward}\ \bibnamefont {Grant}}, \bibinfo {author} {\bibfnamefont {Leonard}\
  \bibnamefont {Wossnig}}, \bibinfo {author} {\bibfnamefont {Ivan}\
  \bibnamefont {Rungger}}, \bibinfo {author} {\bibfnamefont {George~H}\
  \bibnamefont {Booth}},  \emph {et~al.},\ }\bibfield  {title} {\enquote
  {\bibinfo {title} {The variational quantum eigensolver: a review of methods
  and best practices},}\ }\href@noop {} {\bibfield  {journal} {\bibinfo
  {journal} {arXiv preprint arXiv:2111.05176}\ } (\bibinfo {year}
  {2021})}\BibitemShut {NoStop}%
\bibitem [{\citenamefont {Fedorov}\ \emph {et~al.}(2022)\citenamefont
  {Fedorov}, \citenamefont {Peng}, \citenamefont {Govind},\ and\ \citenamefont
  {Alexeev}}]{fedorov2022vqe}%
  \BibitemOpen
  \bibfield  {author} {\bibinfo {author} {\bibfnamefont {Dmitry~A}\
  \bibnamefont {Fedorov}}, \bibinfo {author} {\bibfnamefont {Bo}~\bibnamefont
  {Peng}}, \bibinfo {author} {\bibfnamefont {Niranjan}\ \bibnamefont {Govind}},
  \ and\ \bibinfo {author} {\bibfnamefont {Yuri}\ \bibnamefont {Alexeev}},\
  }\bibfield  {title} {\enquote {\bibinfo {title} {Vqe method: A short survey
  and recent developments},}\ }\href {\doibase 10.1186/s41313-021-00032-6}
  {\bibfield  {journal} {\bibinfo  {journal} {Materials Theory}\ }\textbf
  {\bibinfo {volume} {6}},\ \bibinfo {pages} {1--21} (\bibinfo {year}
  {2022})}\BibitemShut {NoStop}%
\bibitem [{\citenamefont {McClean}\ \emph {et~al.}(2018)\citenamefont
  {McClean}, \citenamefont {Boixo}, \citenamefont {Smelyanskiy}, \citenamefont
  {Babbush},\ and\ \citenamefont {Neven}}]{mcclean2018barren}%
  \BibitemOpen
  \bibfield  {author} {\bibinfo {author} {\bibfnamefont {Jarrod~R}\
  \bibnamefont {McClean}}, \bibinfo {author} {\bibfnamefont {Sergio}\
  \bibnamefont {Boixo}}, \bibinfo {author} {\bibfnamefont {Vadim~N}\
  \bibnamefont {Smelyanskiy}}, \bibinfo {author} {\bibfnamefont {Ryan}\
  \bibnamefont {Babbush}}, \ and\ \bibinfo {author} {\bibfnamefont {Hartmut}\
  \bibnamefont {Neven}},\ }\bibfield  {title} {\enquote {\bibinfo {title}
  {Barren plateaus in quantum neural network training landscapes},}\ }\href
  {\doibase 10.1038/s41467-018-07090-4} {\bibfield  {journal} {\bibinfo
  {journal} {Nature communications}\ }\textbf {\bibinfo {volume} {9}},\
  \bibinfo {pages} {1--6} (\bibinfo {year} {2018})}\BibitemShut {NoStop}%
\bibitem [{\citenamefont {Cade}\ \emph {et~al.}(2020)\citenamefont {Cade},
  \citenamefont {Mineh}, \citenamefont {Montanaro},\ and\ \citenamefont
  {Stanisic}}]{cade2020strategies}%
  \BibitemOpen
  \bibfield  {author} {\bibinfo {author} {\bibfnamefont {Chris}\ \bibnamefont
  {Cade}}, \bibinfo {author} {\bibfnamefont {Lana}\ \bibnamefont {Mineh}},
  \bibinfo {author} {\bibfnamefont {Ashley}\ \bibnamefont {Montanaro}}, \ and\
  \bibinfo {author} {\bibfnamefont {Stasja}\ \bibnamefont {Stanisic}},\
  }\bibfield  {title} {\enquote {\bibinfo {title} {Strategies for solving the
  fermi-hubbard model on near-term quantum computers},}\ }\href {\doibase
  10.1103/PhysRevB.102.235122} {\bibfield  {journal} {\bibinfo  {journal}
  {Physical Review B}\ }\textbf {\bibinfo {volume} {102}},\ \bibinfo {pages}
  {235122} (\bibinfo {year} {2020})}\BibitemShut {NoStop}%
\bibitem [{\citenamefont {Martin}\ \emph {et~al.}(2021)\citenamefont {Martin},
  \citenamefont {Simon},\ and\ \citenamefont
  {Ran{\v{c}}i{\'c}}}]{martin2021simulating}%
  \BibitemOpen
  \bibfield  {author} {\bibinfo {author} {\bibfnamefont {Baptiste~Anselme}\
  \bibnamefont {Martin}}, \bibinfo {author} {\bibfnamefont {Pascal}\
  \bibnamefont {Simon}}, \ and\ \bibinfo {author} {\bibfnamefont {Marko~J}\
  \bibnamefont {Ran{\v{c}}i{\'c}}},\ }\bibfield  {title} {\enquote {\bibinfo
  {title} {Simulating strongly interacting hubbard chains with the variational
  hamiltonian ansatz on a quantum computer},}\ }\href {\doibase
  10.48550/arXiv.2111.11996} {\bibfield  {journal} {\bibinfo  {journal} {arXiv
  preprint arXiv:2111.11996}\ } (\bibinfo {year} {2021}),\
  10.48550/arXiv.2111.11996}\BibitemShut {NoStop}%
\bibitem [{\citenamefont {Kandala}\ \emph {et~al.}(2017)\citenamefont
  {Kandala}, \citenamefont {Mezzacapo}, \citenamefont {Temme}, \citenamefont
  {Takita}, \citenamefont {Brink}, \citenamefont {Chow},\ and\ \citenamefont
  {Gambetta}}]{kandala2017hardware}%
  \BibitemOpen
  \bibfield  {author} {\bibinfo {author} {\bibfnamefont {Abhinav}\ \bibnamefont
  {Kandala}}, \bibinfo {author} {\bibfnamefont {Antonio}\ \bibnamefont
  {Mezzacapo}}, \bibinfo {author} {\bibfnamefont {Kristan}\ \bibnamefont
  {Temme}}, \bibinfo {author} {\bibfnamefont {Maika}\ \bibnamefont {Takita}},
  \bibinfo {author} {\bibfnamefont {Markus}\ \bibnamefont {Brink}}, \bibinfo
  {author} {\bibfnamefont {Jerry~M}\ \bibnamefont {Chow}}, \ and\ \bibinfo
  {author} {\bibfnamefont {Jay~M}\ \bibnamefont {Gambetta}},\ }\bibfield
  {title} {\enquote {\bibinfo {title} {Hardware-efficient variational quantum
  eigensolver for small molecules and quantum magnets},}\ }\href {\doibase
  10.1038/nature23879} {\bibfield  {journal} {\bibinfo  {journal} {Nature}\
  }\textbf {\bibinfo {volume} {549}},\ \bibinfo {pages} {242--246} (\bibinfo
  {year} {2017})}\BibitemShut {NoStop}%
\bibitem [{\citenamefont {Benedetti}\ \emph {et~al.}(2021)\citenamefont
  {Benedetti}, \citenamefont {Fiorentini},\ and\ \citenamefont
  {Lubasch}}]{benedetti2021hardware}%
  \BibitemOpen
  \bibfield  {author} {\bibinfo {author} {\bibfnamefont {Marcello}\
  \bibnamefont {Benedetti}}, \bibinfo {author} {\bibfnamefont {Mattia}\
  \bibnamefont {Fiorentini}}, \ and\ \bibinfo {author} {\bibfnamefont
  {Michael}\ \bibnamefont {Lubasch}},\ }\bibfield  {title} {\enquote {\bibinfo
  {title} {Hardware-efficient variational quantum algorithms for time
  evolution},}\ }\href {\doibase 10.1103/PhysRevResearch.3.033083} {\bibfield
  {journal} {\bibinfo  {journal} {Physical Review Research}\ }\textbf {\bibinfo
  {volume} {3}},\ \bibinfo {pages} {033083} (\bibinfo {year}
  {2021})}\BibitemShut {NoStop}%
\bibitem [{\citenamefont {Rattew}\ \emph {et~al.}(2019)\citenamefont {Rattew},
  \citenamefont {Hu}, \citenamefont {Pistoia}, \citenamefont {Chen},\ and\
  \citenamefont {Wood}}]{rattew2019domain}%
  \BibitemOpen
  \bibfield  {author} {\bibinfo {author} {\bibfnamefont {Arthur~G}\
  \bibnamefont {Rattew}}, \bibinfo {author} {\bibfnamefont {Shaohan}\
  \bibnamefont {Hu}}, \bibinfo {author} {\bibfnamefont {Marco}\ \bibnamefont
  {Pistoia}}, \bibinfo {author} {\bibfnamefont {Richard}\ \bibnamefont {Chen}},
  \ and\ \bibinfo {author} {\bibfnamefont {Steve}\ \bibnamefont {Wood}},\
  }\bibfield  {title} {\enquote {\bibinfo {title} {A domain-agnostic,
  noise-resistant, hardware-efficient evolutionary variational quantum
  eigensolver},}\ }\href {\doibase 10.48550/arXiv.1910.09694} {\bibfield
  {journal} {\bibinfo  {journal} {arXiv preprint arXiv:1910.09694}\ } (\bibinfo
  {year} {2019}),\ 10.48550/arXiv.1910.09694}\BibitemShut {NoStop}%
\bibitem [{\citenamefont {Tang}\ \emph {et~al.}(2021)\citenamefont {Tang},
  \citenamefont {Shkolnikov}, \citenamefont {Barron}, \citenamefont {Grimsley},
  \citenamefont {Mayhall}, \citenamefont {Barnes},\ and\ \citenamefont
  {Economou}}]{tang2021qubit}%
  \BibitemOpen
  \bibfield  {author} {\bibinfo {author} {\bibfnamefont {Ho~Lun}\ \bibnamefont
  {Tang}}, \bibinfo {author} {\bibfnamefont {VO}~\bibnamefont {Shkolnikov}},
  \bibinfo {author} {\bibfnamefont {George~S}\ \bibnamefont {Barron}}, \bibinfo
  {author} {\bibfnamefont {Harper~R}\ \bibnamefont {Grimsley}}, \bibinfo
  {author} {\bibfnamefont {Nicholas~J}\ \bibnamefont {Mayhall}}, \bibinfo
  {author} {\bibfnamefont {Edwin}\ \bibnamefont {Barnes}}, \ and\ \bibinfo
  {author} {\bibfnamefont {Sophia~E}\ \bibnamefont {Economou}},\ }\bibfield
  {title} {\enquote {\bibinfo {title} {qubit-adapt-vqe: An adaptive algorithm
  for constructing hardware-efficient ans{\"a}tze on a quantum processor},}\
  }\href {\doibase 10.1103/PRXQuantum.2.020310} {\bibfield  {journal} {\bibinfo
   {journal} {PRX Quantum}\ }\textbf {\bibinfo {volume} {2}},\ \bibinfo {pages}
  {020310} (\bibinfo {year} {2021})}\BibitemShut {NoStop}%
\bibitem [{\citenamefont {Sim}\ \emph {et~al.}(2021)\citenamefont {Sim},
  \citenamefont {Romero}, \citenamefont {Gonthier},\ and\ \citenamefont
  {Kunitsa}}]{sim2021adaptive}%
  \BibitemOpen
  \bibfield  {author} {\bibinfo {author} {\bibfnamefont {Sukin}\ \bibnamefont
  {Sim}}, \bibinfo {author} {\bibfnamefont {Jonathan}\ \bibnamefont {Romero}},
  \bibinfo {author} {\bibfnamefont {J{\'e}r{\^o}me~F}\ \bibnamefont
  {Gonthier}}, \ and\ \bibinfo {author} {\bibfnamefont {Alexander~A}\
  \bibnamefont {Kunitsa}},\ }\bibfield  {title} {\enquote {\bibinfo {title}
  {Adaptive pruning-based optimization of parameterized quantum circuits},}\
  }\href {\doibase 10.1088/2058-9565/abe107} {\bibfield  {journal} {\bibinfo
  {journal} {Quantum Science and Technology}\ }\textbf {\bibinfo {volume}
  {6}},\ \bibinfo {pages} {025019} (\bibinfo {year} {2021})}\BibitemShut
  {NoStop}%
\bibitem [{\citenamefont {Chivilikhin}\ \emph {et~al.}(2020)\citenamefont
  {Chivilikhin}, \citenamefont {Samarin}, \citenamefont {Ulyantsev},
  \citenamefont {Iorsh}, \citenamefont {Oganov},\ and\ \citenamefont
  {Kyriienko}}]{chivilikhin2020mog}%
  \BibitemOpen
  \bibfield  {author} {\bibinfo {author} {\bibfnamefont {D}~\bibnamefont
  {Chivilikhin}}, \bibinfo {author} {\bibfnamefont {A}~\bibnamefont {Samarin}},
  \bibinfo {author} {\bibfnamefont {V}~\bibnamefont {Ulyantsev}}, \bibinfo
  {author} {\bibfnamefont {I}~\bibnamefont {Iorsh}}, \bibinfo {author}
  {\bibfnamefont {AR}~\bibnamefont {Oganov}}, \ and\ \bibinfo {author}
  {\bibfnamefont {O}~\bibnamefont {Kyriienko}},\ }\bibfield  {title} {\enquote
  {\bibinfo {title} {Mog-vqe: Multiobjective genetic variational quantum
  eigensolver},}\ }\href {\doibase 10.48550/arXiv.2007.04424} {\bibfield
  {journal} {\bibinfo  {journal} {arXiv preprint arXiv:2007.04424}\ } (\bibinfo
  {year} {2020}),\ 10.48550/arXiv.2007.04424}\BibitemShut {NoStop}%
\bibitem [{\citenamefont {Meister}\ \emph {et~al.}(2022)\citenamefont
  {Meister}, \citenamefont {Gustiani},\ and\ \citenamefont
  {Benjamin}}]{sister}%
  \BibitemOpen
  \bibfield  {author} {\bibinfo {author} {\bibfnamefont {Richard}\ \bibnamefont
  {Meister}}, \bibinfo {author} {\bibfnamefont {Cica}\ \bibnamefont
  {Gustiani}}, \ and\ \bibinfo {author} {\bibfnamefont {Simon~C.}\ \bibnamefont
  {Benjamin}},\ }\bibfield  {title} {\enquote {\bibinfo {title} {Exploring ab
  initio machine synthesis of quantum circuits},}\ }\href@noop {} {\bibfield
  {journal} {\bibinfo  {journal} {in preparation}\ } (\bibinfo {year}
  {2022})}\BibitemShut {NoStop}%
\bibitem [{\citenamefont {Szabo}\ and\ \citenamefont
  {Ostlund}(2012)}]{szabo2012modern}%
  \BibitemOpen
  \bibfield  {author} {\bibinfo {author} {\bibfnamefont {Attila}\ \bibnamefont
  {Szabo}}\ and\ \bibinfo {author} {\bibfnamefont {Neil~S}\ \bibnamefont
  {Ostlund}},\ }\href@noop {} {\emph {\bibinfo {title} {Modern quantum
  chemistry: introduction to advanced electronic structure theory}}}\ (\bibinfo
   {publisher} {Courier Corporation},\ \bibinfo {year} {2012})\BibitemShut
  {NoStop}%
\bibitem [{\citenamefont {Wigner}\ and\ \citenamefont
  {Jordan}(1928)}]{wigner1928paulische}%
  \BibitemOpen
  \bibfield  {author} {\bibinfo {author} {\bibfnamefont {E}~\bibnamefont
  {Wigner}}\ and\ \bibinfo {author} {\bibfnamefont {Pascual}\ \bibnamefont
  {Jordan}},\ }\bibfield  {title} {\enquote {\bibinfo {title} {{\"U}ber das
  paulische {\"a}quivalenzverbot},}\ }\href {\doibase 10.1007/BF01331938}
  {\bibfield  {journal} {\bibinfo  {journal} {Z. Phys}\ }\textbf {\bibinfo
  {volume} {47}},\ \bibinfo {pages} {631} (\bibinfo {year} {1928})}\BibitemShut
  {NoStop}%
\bibitem [{\citenamefont {Seeley}\ \emph {et~al.}(2012)\citenamefont {Seeley},
  \citenamefont {Richard},\ and\ \citenamefont {Love}}]{seeley2012bravyi}%
  \BibitemOpen
  \bibfield  {author} {\bibinfo {author} {\bibfnamefont {Jacob~T}\ \bibnamefont
  {Seeley}}, \bibinfo {author} {\bibfnamefont {Martin~J}\ \bibnamefont
  {Richard}}, \ and\ \bibinfo {author} {\bibfnamefont {Peter~J}\ \bibnamefont
  {Love}},\ }\bibfield  {title} {\enquote {\bibinfo {title} {The bravyi-kitaev
  transformation for quantum computation of electronic structure},}\ }\href
  {\doibase 10.1063/1.4768229} {\bibfield  {journal} {\bibinfo  {journal} {The
  Journal of chemical physics}\ }\textbf {\bibinfo {volume} {137}},\ \bibinfo
  {pages} {224109} (\bibinfo {year} {2012})}\BibitemShut {NoStop}%
\bibitem [{\citenamefont {McClean}\ \emph {et~al.}(2017)\citenamefont
  {McClean}, \citenamefont {Kivlichan}, \citenamefont {Sung}, \citenamefont
  {Steiger}, \citenamefont {Cao}, \citenamefont {Dai}, \citenamefont {Fried},
  \citenamefont {Gidney}, \citenamefont {Gimby}, \citenamefont {Gokhale} \emph
  {et~al.}}]{mcclean2017openfermion}%
  \BibitemOpen
  \bibfield  {author} {\bibinfo {author} {\bibfnamefont {JR}~\bibnamefont
  {McClean}}, \bibinfo {author} {\bibfnamefont {ID}~\bibnamefont {Kivlichan}},
  \bibinfo {author} {\bibfnamefont {KJ}~\bibnamefont {Sung}}, \bibinfo {author}
  {\bibfnamefont {DS}~\bibnamefont {Steiger}}, \bibinfo {author} {\bibfnamefont
  {Y}~\bibnamefont {Cao}}, \bibinfo {author} {\bibfnamefont {C}~\bibnamefont
  {Dai}}, \bibinfo {author} {\bibfnamefont {E~Schuyler}\ \bibnamefont {Fried}},
  \bibinfo {author} {\bibfnamefont {C}~\bibnamefont {Gidney}}, \bibinfo
  {author} {\bibfnamefont {B}~\bibnamefont {Gimby}}, \bibinfo {author}
  {\bibfnamefont {P}~\bibnamefont {Gokhale}},  \emph {et~al.},\ }\bibfield
  {title} {\enquote {\bibinfo {title} {Openfermion: The electronic structure
  package for quantum computers.}}\ }\href {\doibase 10.48550/arXiv.1710.07629}
  {\bibfield  {journal} {\bibinfo  {journal} {arXiv preprint
  quant-ph/1710.07629}\ } (\bibinfo {year} {2017}),\
  10.48550/arXiv.1710.07629}\BibitemShut {NoStop}%
\bibitem [{\citenamefont {Sun}\ \emph {et~al.}(2018)\citenamefont {Sun},
  \citenamefont {Berkelbach}, \citenamefont {Blunt}, \citenamefont {Booth},
  \citenamefont {Guo}, \citenamefont {Li}, \citenamefont {Liu}, \citenamefont
  {McClain}, \citenamefont {Sayfutyarova}, \citenamefont {Sharma} \emph
  {et~al.}}]{sun2018pyscf}%
  \BibitemOpen
  \bibfield  {author} {\bibinfo {author} {\bibfnamefont {Qiming}\ \bibnamefont
  {Sun}}, \bibinfo {author} {\bibfnamefont {Timothy~C}\ \bibnamefont
  {Berkelbach}}, \bibinfo {author} {\bibfnamefont {Nick~S}\ \bibnamefont
  {Blunt}}, \bibinfo {author} {\bibfnamefont {George~H}\ \bibnamefont {Booth}},
  \bibinfo {author} {\bibfnamefont {Sheng}\ \bibnamefont {Guo}}, \bibinfo
  {author} {\bibfnamefont {Zhendong}\ \bibnamefont {Li}}, \bibinfo {author}
  {\bibfnamefont {Junzi}\ \bibnamefont {Liu}}, \bibinfo {author} {\bibfnamefont
  {James~D}\ \bibnamefont {McClain}}, \bibinfo {author} {\bibfnamefont
  {Elvira~R}\ \bibnamefont {Sayfutyarova}}, \bibinfo {author} {\bibfnamefont
  {Sandeep}\ \bibnamefont {Sharma}},  \emph {et~al.},\ }\bibfield  {title}
  {\enquote {\bibinfo {title} {Pyscf: the python-based simulations of chemistry
  framework},}\ }\href {\doibase 10.1002/wcms.1340} {\bibfield  {journal}
  {\bibinfo  {journal} {Wiley Interdisciplinary Reviews: Computational
  Molecular Science}\ }\textbf {\bibinfo {volume} {8}},\ \bibinfo {pages}
  {e1340} (\bibinfo {year} {2018})}\BibitemShut {NoStop}%
\bibitem [{\citenamefont {Pople}(1999)}]{pople1999nobel}%
  \BibitemOpen
  \bibfield  {author} {\bibinfo {author} {\bibfnamefont {John~A}\ \bibnamefont
  {Pople}},\ }\bibfield  {title} {\enquote {\bibinfo {title} {Nobel lecture:
  Quantum chemical models},}\ }\href {\doibase 10.1103/RevModPhys.71.1267}
  {\bibfield  {journal} {\bibinfo  {journal} {Reviews of Modern Physics}\
  }\textbf {\bibinfo {volume} {71}},\ \bibinfo {pages} {1267} (\bibinfo {year}
  {1999})}\BibitemShut {NoStop}%
\bibitem [{\citenamefont {Hatano}\ and\ \citenamefont
  {Suzuki}(2005)}]{trotter}%
  \BibitemOpen
  \bibfield  {author} {\bibinfo {author} {\bibfnamefont {Naomichi}\
  \bibnamefont {Hatano}}\ and\ \bibinfo {author} {\bibfnamefont {Masuo}\
  \bibnamefont {Suzuki}},\ }\bibfield  {title} {\enquote {\bibinfo {title}
  {Quantum annealing and related optimization methods},}\ }\href@noop {}
  {\bibfield  {journal} {\bibinfo  {journal} {Quantum Annealing and Related
  Optimization Methods, Edited by A. Das and B.K. Chakrabarti. 2005 XIV, 378 p.
  124 illus. Also available online. ISBN 3-540-27987-3. Berlin: Springer,
  2005.}\ } (\bibinfo {year} {2005})}\BibitemShut {NoStop}%
\bibitem [{\citenamefont {Sharma}\ \emph {et~al.}(2020)\citenamefont {Sharma},
  \citenamefont {Khatri}, \citenamefont {Cerezo},\ and\ \citenamefont
  {Coles}}]{sharma2020noise}%
  \BibitemOpen
  \bibfield  {author} {\bibinfo {author} {\bibfnamefont {Kunal}\ \bibnamefont
  {Sharma}}, \bibinfo {author} {\bibfnamefont {Sumeet}\ \bibnamefont {Khatri}},
  \bibinfo {author} {\bibfnamefont {Marco}\ \bibnamefont {Cerezo}}, \ and\
  \bibinfo {author} {\bibfnamefont {Patrick~J}\ \bibnamefont {Coles}},\
  }\bibfield  {title} {\enquote {\bibinfo {title} {Noise resilience of
  variational quantum compiling},}\ }\href {\doibase 10.1088/1367-2630/ab784c}
  {\bibfield  {journal} {\bibinfo  {journal} {New Journal of Physics}\ }\textbf
  {\bibinfo {volume} {22}},\ \bibinfo {pages} {043006} (\bibinfo {year}
  {2020})}\BibitemShut {NoStop}%
\bibitem [{\citenamefont {McArdle}\ \emph {et~al.}(2019)\citenamefont
  {McArdle}, \citenamefont {Jones}, \citenamefont {Endo}, \citenamefont {Li},
  \citenamefont {Benjamin},\ and\ \citenamefont
  {Yuan}}]{mcardle2019variational}%
  \BibitemOpen
  \bibfield  {author} {\bibinfo {author} {\bibfnamefont {Sam}\ \bibnamefont
  {McArdle}}, \bibinfo {author} {\bibfnamefont {Tyson}\ \bibnamefont {Jones}},
  \bibinfo {author} {\bibfnamefont {Suguru}\ \bibnamefont {Endo}}, \bibinfo
  {author} {\bibfnamefont {Ying}\ \bibnamefont {Li}}, \bibinfo {author}
  {\bibfnamefont {Simon~C}\ \bibnamefont {Benjamin}}, \ and\ \bibinfo {author}
  {\bibfnamefont {Xiao}\ \bibnamefont {Yuan}},\ }\bibfield  {title} {\enquote
  {\bibinfo {title} {Variational ansatz-based quantum simulation of imaginary
  time evolution},}\ }\href {\doibase 10.1038/s41534-019-0187-2} {\bibfield
  {journal} {\bibinfo  {journal} {npj Quantum Information}\ }\textbf {\bibinfo
  {volume} {5}},\ \bibinfo {pages} {1--6} (\bibinfo {year} {2019})}\BibitemShut
  {NoStop}%
\bibitem [{\citenamefont {Stokes}\ \emph {et~al.}(2020)\citenamefont {Stokes},
  \citenamefont {Izaac}, \citenamefont {Killoran},\ and\ \citenamefont
  {Carleo}}]{stokes2020quantum}%
  \BibitemOpen
  \bibfield  {author} {\bibinfo {author} {\bibfnamefont {James}\ \bibnamefont
  {Stokes}}, \bibinfo {author} {\bibfnamefont {Josh}\ \bibnamefont {Izaac}},
  \bibinfo {author} {\bibfnamefont {Nathan}\ \bibnamefont {Killoran}}, \ and\
  \bibinfo {author} {\bibfnamefont {Giuseppe}\ \bibnamefont {Carleo}},\
  }\bibfield  {title} {\enquote {\bibinfo {title} {Quantum natural gradient},}\
  }\href {\doibase 10.22331/q-2020-05-25-269} {\bibfield  {journal} {\bibinfo
  {journal} {Quantum}\ }\textbf {\bibinfo {volume} {4}},\ \bibinfo {pages}
  {269} (\bibinfo {year} {2020})}\BibitemShut {NoStop}%
\bibitem [{\citenamefont {Meyer}(2021)}]{meyer2021fisher}%
  \BibitemOpen
  \bibfield  {author} {\bibinfo {author} {\bibfnamefont {Johannes~Jakob}\
  \bibnamefont {Meyer}},\ }\bibfield  {title} {\enquote {\bibinfo {title}
  {Fisher information in noisy intermediate-scale quantum applications},}\
  }\href {\doibase 10.22331/q-2021-09-09-539} {\bibfield  {journal} {\bibinfo
  {journal} {Quantum}\ }\textbf {\bibinfo {volume} {5}},\ \bibinfo {pages}
  {539} (\bibinfo {year} {2021})}\BibitemShut {NoStop}%
\bibitem [{\citenamefont {Gustiani}(2022)}]{cicacica}%
  \BibitemOpen
  \bibfield  {author} {\bibinfo {author} {\bibfnamefont {Cica}\ \bibnamefont
  {Gustiani}},\ }\href {https://github.com/cicacica/chemistry-code.git}
  {\enquote {\bibinfo {title} {Supplementary materials of paper entitled
  exploiting subspace constraints and ab initio variational methods for quantum
  chemistry},}\ } (\bibinfo {year} {2022})\BibitemShut {NoStop}%
\bibitem [{\citenamefont {Jones}\ and\ \citenamefont
  {Benjamin}(2020)}]{jones2020questlink}%
  \BibitemOpen
  \bibfield  {author} {\bibinfo {author} {\bibfnamefont {Tyson}\ \bibnamefont
  {Jones}}\ and\ \bibinfo {author} {\bibfnamefont {Simon}\ \bibnamefont
  {Benjamin}},\ }\bibfield  {title} {\enquote {\bibinfo {title}
  {Questlink—mathematica embiggened by a hardware-optimised quantum
  emulator},}\ }\href {\doibase 10.1088/2058-9565/ab8506} {\bibfield  {journal}
  {\bibinfo  {journal} {Quantum Science and Technology}\ }\textbf {\bibinfo
  {volume} {5}},\ \bibinfo {pages} {034012} (\bibinfo {year}
  {2020})}\BibitemShut {NoStop}%
\bibitem [{\citenamefont {Jones}\ \emph {et~al.}(2019)\citenamefont {Jones},
  \citenamefont {Brown}, \citenamefont {Bush},\ and\ \citenamefont
  {Benjamin}}]{jones2019quest}%
  \BibitemOpen
  \bibfield  {author} {\bibinfo {author} {\bibfnamefont {Tyson}\ \bibnamefont
  {Jones}}, \bibinfo {author} {\bibfnamefont {Anna}\ \bibnamefont {Brown}},
  \bibinfo {author} {\bibfnamefont {Ian}\ \bibnamefont {Bush}}, \ and\ \bibinfo
  {author} {\bibfnamefont {Simon~C}\ \bibnamefont {Benjamin}},\ }\bibfield
  {title} {\enquote {\bibinfo {title} {Quest and high performance simulation of
  quantum computers},}\ }\href {\doibase 10.1038/s41598-019-47174-9} {\bibfield
   {journal} {\bibinfo  {journal} {Scientific reports}\ }\textbf {\bibinfo
  {volume} {9}},\ \bibinfo {pages} {1--11} (\bibinfo {year}
  {2019})}\BibitemShut {NoStop}%
\bibitem [{\citenamefont {Meister}(2022)}]{pyquest}%
  \BibitemOpen
  \bibfield  {author} {\bibinfo {author} {\bibfnamefont {Richard}\ \bibnamefont
  {Meister}},\ }\href {https://github.com/rrmeister/pyQuEST} {\enquote
  {\bibinfo {title} {\texttt{pyQuEST}},}\ } (\bibinfo {year}
  {2022})\BibitemShut {NoStop}%
\bibitem [{\citenamefont {Claudino}\ \emph {et~al.}(2020)\citenamefont
  {Claudino}, \citenamefont {Wright}, \citenamefont {McCaskey},\ and\
  \citenamefont {Humble}}]{claudino2020benchmarking}%
  \BibitemOpen
  \bibfield  {author} {\bibinfo {author} {\bibfnamefont {Daniel}\ \bibnamefont
  {Claudino}}, \bibinfo {author} {\bibfnamefont {Jerimiah}\ \bibnamefont
  {Wright}}, \bibinfo {author} {\bibfnamefont {Alexander~J}\ \bibnamefont
  {McCaskey}}, \ and\ \bibinfo {author} {\bibfnamefont {Travis~S}\ \bibnamefont
  {Humble}},\ }\bibfield  {title} {\enquote {\bibinfo {title} {Benchmarking
  adaptive variational quantum eigensolvers},}\ }\href {\doibase
  10.3389/fchem.2020.606863} {\bibfield  {journal} {\bibinfo  {journal}
  {Frontiers in Chemistry}\ }\textbf {\bibinfo {volume} {8}},\ \bibinfo {pages}
  {1152} (\bibinfo {year} {2020})}\BibitemShut {NoStop}%
\bibitem [{\citenamefont {Tkachenko}\ \emph {et~al.}(2021)\citenamefont
  {Tkachenko}, \citenamefont {Sud}, \citenamefont {Zhang}, \citenamefont
  {Tretiak}, \citenamefont {Anisimov}, \citenamefont {Arrasmith}, \citenamefont
  {Coles}, \citenamefont {Cincio},\ and\ \citenamefont
  {Dub}}]{tkachenko2021correlation}%
  \BibitemOpen
  \bibfield  {author} {\bibinfo {author} {\bibfnamefont {Nikolay~V}\
  \bibnamefont {Tkachenko}}, \bibinfo {author} {\bibfnamefont {James}\
  \bibnamefont {Sud}}, \bibinfo {author} {\bibfnamefont {Yu}~\bibnamefont
  {Zhang}}, \bibinfo {author} {\bibfnamefont {Sergei}\ \bibnamefont {Tretiak}},
  \bibinfo {author} {\bibfnamefont {Petr~M}\ \bibnamefont {Anisimov}}, \bibinfo
  {author} {\bibfnamefont {Andrew~T}\ \bibnamefont {Arrasmith}}, \bibinfo
  {author} {\bibfnamefont {Patrick~J}\ \bibnamefont {Coles}}, \bibinfo {author}
  {\bibfnamefont {Lukasz}\ \bibnamefont {Cincio}}, \ and\ \bibinfo {author}
  {\bibfnamefont {Pavel~A}\ \bibnamefont {Dub}},\ }\bibfield  {title} {\enquote
  {\bibinfo {title} {Correlation-informed permutation of qubits for reducing
  ansatz depth in the variational quantum eigensolver},}\ }\href {\doibase
  10.1103/PRXQuantum.2.020337} {\bibfield  {journal} {\bibinfo  {journal} {PRX
  Quantum}\ }\textbf {\bibinfo {volume} {2}},\ \bibinfo {pages} {020337}
  (\bibinfo {year} {2021})}\BibitemShut {NoStop}%
\bibitem [{\citenamefont {Gomes}\ \emph {et~al.}(2021)\citenamefont {Gomes},
  \citenamefont {Mukherjee}, \citenamefont {Zhang}, \citenamefont {Iadecola},
  \citenamefont {Wang}, \citenamefont {Ho}, \citenamefont {Orth},\ and\
  \citenamefont {Yao}}]{gomes2021adaptive}%
  \BibitemOpen
  \bibfield  {author} {\bibinfo {author} {\bibfnamefont {Niladri}\ \bibnamefont
  {Gomes}}, \bibinfo {author} {\bibfnamefont {Anirban}\ \bibnamefont
  {Mukherjee}}, \bibinfo {author} {\bibfnamefont {Feng}\ \bibnamefont {Zhang}},
  \bibinfo {author} {\bibfnamefont {Thomas}\ \bibnamefont {Iadecola}}, \bibinfo
  {author} {\bibfnamefont {Cai-Zhuang}\ \bibnamefont {Wang}}, \bibinfo {author}
  {\bibfnamefont {Kai-Ming}\ \bibnamefont {Ho}}, \bibinfo {author}
  {\bibfnamefont {Peter~P.}\ \bibnamefont {Orth}}, \ and\ \bibinfo {author}
  {\bibfnamefont {Yong-Xin}\ \bibnamefont {Yao}},\ }\bibfield  {title}
  {\enquote {\bibinfo {title} {Adaptive variational quantum imaginary time
  evolution approach for ground state preparation},}\ }\href {\doibase
  10.1002/qute.202100114} {\bibfield  {journal} {\bibinfo  {journal} {Advanced
  Quantum Technologies}\ }\textbf {\bibinfo {volume} {4}},\ \bibinfo {pages}
  {2100114} (\bibinfo {year} {2021})}\BibitemShut {NoStop}%
\bibitem [{\citenamefont {Bravyi}\ \emph {et~al.}(2017)\citenamefont {Bravyi},
  \citenamefont {Gambetta}, \citenamefont {Mezzacapo},\ and\ \citenamefont
  {Temme}}]{bravyi2017tapering}%
  \BibitemOpen
  \bibfield  {author} {\bibinfo {author} {\bibfnamefont {Sergey}\ \bibnamefont
  {Bravyi}}, \bibinfo {author} {\bibfnamefont {Jay~M}\ \bibnamefont
  {Gambetta}}, \bibinfo {author} {\bibfnamefont {Antonio}\ \bibnamefont
  {Mezzacapo}}, \ and\ \bibinfo {author} {\bibfnamefont {Kristan}\ \bibnamefont
  {Temme}},\ }\bibfield  {title} {\enquote {\bibinfo {title} {Tapering off
  qubits to simulate fermionic hamiltonians},}\ }\href {\doibase
  10.48550/arXiv.1701.08213} {\bibfield  {journal} {\bibinfo  {journal} {arXiv
  preprint arXiv:1701.08213}\ } (\bibinfo {year} {2017}),\
  10.48550/arXiv.1701.08213}\BibitemShut {NoStop}%
\bibitem [{\citenamefont {Jaeger}(2007)}]{jaeger2007quantum}%
  \BibitemOpen
  \bibfield  {author} {\bibinfo {author} {\bibfnamefont {Gregg}\ \bibnamefont
  {Jaeger}},\ }\href {\doibase 10.1007/978-0-387-36944-0} {\emph {\bibinfo
  {title} {Quantum Information: An Overview}}}\ (\bibinfo  {publisher}
  {Springer New York},\ \bibinfo {year} {2007})\BibitemShut {NoStop}%
\bibitem [{\citenamefont {Barenco}\ \emph {et~al.}(1995)\citenamefont
  {Barenco}, \citenamefont {Bennett}, \citenamefont {Cleve}, \citenamefont
  {DiVincenzo}, \citenamefont {Margolus}, \citenamefont {Shor}, \citenamefont
  {Sleator}, \citenamefont {Smolin},\ and\ \citenamefont
  {Weinfurter}}]{barenco1995elementary}%
  \BibitemOpen
  \bibfield  {author} {\bibinfo {author} {\bibfnamefont {Adriano}\ \bibnamefont
  {Barenco}}, \bibinfo {author} {\bibfnamefont {Charles~H}\ \bibnamefont
  {Bennett}}, \bibinfo {author} {\bibfnamefont {Richard}\ \bibnamefont
  {Cleve}}, \bibinfo {author} {\bibfnamefont {David~P}\ \bibnamefont
  {DiVincenzo}}, \bibinfo {author} {\bibfnamefont {Norman}\ \bibnamefont
  {Margolus}}, \bibinfo {author} {\bibfnamefont {Peter}\ \bibnamefont {Shor}},
  \bibinfo {author} {\bibfnamefont {Tycho}\ \bibnamefont {Sleator}}, \bibinfo
  {author} {\bibfnamefont {John~A}\ \bibnamefont {Smolin}}, \ and\ \bibinfo
  {author} {\bibfnamefont {Harald}\ \bibnamefont {Weinfurter}},\ }\bibfield
  {title} {\enquote {\bibinfo {title} {Elementary gates for quantum
  computation},}\ }\href {\doibase 10.1103/PhysRevA.52.3457} {\bibfield
  {journal} {\bibinfo  {journal} {Physical review A}\ }\textbf {\bibinfo
  {volume} {52}},\ \bibinfo {pages} {3457} (\bibinfo {year}
  {1995})}\BibitemShut {NoStop}%
\bibitem [{\citenamefont {Fleck}\ \emph {et~al.}(1976)\citenamefont {Fleck},
  \citenamefont {Morris},\ and\ \citenamefont {Feit}}]{fleck1976time}%
  \BibitemOpen
  \bibfield  {author} {\bibinfo {author} {\bibfnamefont {Jr~A}\ \bibnamefont
  {Fleck}}, \bibinfo {author} {\bibfnamefont {JR}~\bibnamefont {Morris}}, \
  and\ \bibinfo {author} {\bibfnamefont {MD}~\bibnamefont {Feit}},\ }\bibfield
  {title} {\enquote {\bibinfo {title} {Time-dependent propagation of high
  energy laser beams through the atmosphere},}\ }\href {\doibase
  10.1007/BF00896333} {\bibfield  {journal} {\bibinfo  {journal} {Applied
  physics}\ }\textbf {\bibinfo {volume} {10}},\ \bibinfo {pages} {129--160}
  (\bibinfo {year} {1976})}\BibitemShut {NoStop}%
\bibitem [{\citenamefont {Kassal}\ \emph {et~al.}(2008)\citenamefont {Kassal},
  \citenamefont {Jordan}, \citenamefont {Love}, \citenamefont {Mohseni},\ and\
  \citenamefont {Aspuru-Guzik}}]{kassal2008polynomial}%
  \BibitemOpen
  \bibfield  {author} {\bibinfo {author} {\bibfnamefont {Ivan}\ \bibnamefont
  {Kassal}}, \bibinfo {author} {\bibfnamefont {Stephen~P}\ \bibnamefont
  {Jordan}}, \bibinfo {author} {\bibfnamefont {Peter~J}\ \bibnamefont {Love}},
  \bibinfo {author} {\bibfnamefont {Masoud}\ \bibnamefont {Mohseni}}, \ and\
  \bibinfo {author} {\bibfnamefont {Al{\'a}n}\ \bibnamefont {Aspuru-Guzik}},\
  }\bibfield  {title} {\enquote {\bibinfo {title} {Polynomial-time quantum
  algorithm for the simulation of chemical dynamics},}\ }\href {\doibase
  10.1073/pnas.0808245105} {\bibfield  {journal} {\bibinfo  {journal}
  {Proceedings of the National Academy of Sciences}\ }\textbf {\bibinfo
  {volume} {105}},\ \bibinfo {pages} {18681--18686} (\bibinfo {year}
  {2008})}\BibitemShut {NoStop}%
\bibitem [{\citenamefont {Chan}\ \emph {et~al.}(2022)\citenamefont {Chan},
  \citenamefont {Meister}, \citenamefont {Jones}, \citenamefont {Tew},\ and\
  \citenamefont {Benjamin}}]{chan2022grid}%
  \BibitemOpen
  \bibfield  {author} {\bibinfo {author} {\bibfnamefont {Hans Hon~Sang}\
  \bibnamefont {Chan}}, \bibinfo {author} {\bibfnamefont {Richard}\
  \bibnamefont {Meister}}, \bibinfo {author} {\bibfnamefont {Tyson}\
  \bibnamefont {Jones}}, \bibinfo {author} {\bibfnamefont {David~P}\
  \bibnamefont {Tew}}, \ and\ \bibinfo {author} {\bibfnamefont {Simon~C}\
  \bibnamefont {Benjamin}},\ }\bibfield  {title} {\enquote {\bibinfo {title}
  {Grid-based methods for chemistry modelling on a quantum computer},}\ }\href
  {\doibase 10.48550/arXiv.2202.05864} {\bibfield  {journal} {\bibinfo
  {journal} {arXiv preprint arXiv:2202.05864}\ } (\bibinfo {year} {2022}),\
  10.48550/arXiv.2202.05864}\BibitemShut {NoStop}%
\end{thebibliography}%
